\documentclass[usenatbib]{mn2e}
\usepackage{epsfig,lscape,color,graphicx}

\newcommand{\etal}{et~al.}

\newcommand{\kms}{$\mbox{km~s}^{-1}$ }
\newcommand{\kmsns}{$\mbox{km~s}^{-1}$}

\newcommand{\twofig}[2]        
{
   \begin{center}
     \begin{minipage}[t]{0.5\textwidth}
         \psfig{file=eps/#1.eps,height=0.95\textwidth}
     \end{minipage}
     \hfill
     \begin{minipage}[t]{0.5\textwidth}
         \psfig{file=eps/#2.eps,height=0.95\textwidth}
     \end{minipage}
   \end{center}
}

\begin{document}

\title[MMB versus HOPS]{The 6-GHz multibeam maser survey III: comparison between the MMB and HOPS}
\author[S.\ L.\ Breen \etal]{S.\ L. Breen,$^{1}$\thanks{Email: Shari.Breen@sydney.edu.au}, Y. Contreras,$^2$ S.\ P. Ellingsen,$^3$ J.\ A. Green,$^4$ A.\ J. Walsh,$^5$ A. Avison,$^{6,7}$  \newauthor S.\ N. Longmore,$^8$ G.\ A. Fuller,$^{6,7}$ M. A. Voronkov,$^4$  J. Horton,$^1$ A. Kroon$^1$ \\ 
 \\
  $^1$ Sydney Institute for Astronomy (SIfA), School of Physics, University of Sydney, NSW 2006, Australia;\\
  $^2$ Leiden Observatory, Leiden University, P.O. Box 9513, NL-2300 RA Leiden, The Netherlands;\\
  $^3$ School of Mathematics and Physics, University of Tasmania, Private Bag 37, Hobart, Tasmania 7001, Australia;\\
  $^4$ CSIRO Astronomy and Space Science, Australia Telescope National Facility, PO Box 76, Epping, NSW 1710, Australia;\\
  $^5$ International Centre for Radio Astronomy Research, Curtin University, GPO Box U1987, Perth, WA 6845, Australia;\\
  $^6$ Jodrell Bank Centre for Astrophysics, Alan Turing Building, School of Physics and Astronomy, The University of Manchester,\\ Manchester M13 9PL, UK;\\
  $^7$ UK ALMA Regional Centre Node, Manchester, M13 9PL, UK;\\
  $^8$ Astrophysics Research Institute, Liverpool John Moores University, 146 Brownlow Hill, Liverpool L3 5RF, UK\\}


 \maketitle
  
 \begin{abstract}
We have compared the occurrence of 6.7-GHz and 12.2-GHz methanol masers with 22-GHz water masers and 6035-MHz excited-state OH masers in the 100 square degree region of the southern Galactic plane common to the Methanol Multibeam (MMB) and H$_2$O southern Galactic Plane surveys (HOPS). We find the most populous star formation species to be 6.7-GHz methanol, followed by water, then 12.2-GHz and, finally, excited-state OH masers. We present association statistics, flux density (and luminosity where appropriate) and velocity range distributions across the largest, fully surveyed portion of the Galactic plane for four of the most common types of masers found in the vicinity of star formation regions. 

Comparison of the occurrence of the four maser types with far-infrared dust temperatures shows that sources exhibiting excited-state OH maser emission are warmer than sources showing any of the other three maser types. We further find that sources exhibiting both 6.7-GHz and 12.2-GHz methanol masers are warmer than sources exhibiting just 6.7-GHz methanol maser emission. These findings are consistent with previously made suggestions that both OH and 12.2-GHz methanol masers generally trace a later stage of star formation compared to other common maser types.

\end{abstract}


\section{Introduction}

Masers are important probes of a number of types of astronomical objects, and are particularly prevalent towards regions of high-mass star formation. Two of the most commonly detected masers in our Galaxy arise from the 6.7-GHz methanol and 22-GHz water transitions \citep[e.g.][]{Breen15,Walsh14}, followed by a host of other transitions that include the 12.2-GHz methanol maser \citep[e.g.][]{Breen16} and 6035-MHz excited OH masers \citep[e.g.][]{Avison16}. Other, rarer, maser transitions also have a special role to play - signposting physical conditions that are less commonly found, or, perhaps more likely, associated with short-lived evolutionary phases in the star formation process \citep[e.g.][]{Ellingsen11,Ellingsen13}.  

The exclusive association between 6.7-GHz methanol masers and young high-mass stars \citep[e.g.][]{Minier03,Xu08,Breen13} make them particularly useful for pinpointing and studying high-mass star formation. Since they also have a tendency to trace systemic velocities \citep[e.g.][]{Szy07,C09,Pandian09,GM11} they have been successfully used to trace Galactic structure \citep{Green11}. They are relatively common and strong, with more than 1000 sources now known across the Galactic plane \citep[e.g.][]{Pandian07,CasMMB10,GreenMMB10,CasMMB102,Green12,Breen15}. The next strongest and common class II methanol maser line is at 12.2-GHz \citep[e.g.][]{Caswell95b,Blas04,Gay,BreenMMB12a,BreenMMB12b,BreenMMB14,Breen16}. 

Methanol masers at 12.2-GHz have always been found to have 6.7-GHz counterparts, and although a large, sensitive, unbiased survey for 12.2-GHz methanol maser emission has never been conducted, we can be relatively certain that a near-complete sample of these masers can be gained by targeting a complete sample of 6.7-GHz masers. While in recent years most 12.2-GHz methanol maser searches have been targeted towards 6.7-GHz methanol masers, some of the earlier searches targeted known sites of OH maser emission \citep[e.g.][]{Kemball88,Cas93}, water maser emission \citep[e.g.][]{Koo88} and other sources indicative of star formation regions \citep[e.g.][]{Norris1987}. All of these early 12.2-GHz detections were later reported as sources of 6.7-GHz methanol maser emission, and this, combined with the fact that there has never been a serendipitous detection of 12.2-GHz methanol maser emission without a 6.7-GHz counterpart, and the fact that 12.2-GHz methanol maser emission only rarely shows peak flux densities that surpass that of their 6.7-GHz counterpart \citep[e.g.][]{Breen12stats} clearly indicates that there is not a significant population of 12.2-GHz methanol masers devoid of 6.7-GHz counterparts.

The two transitions are typically co-spatial to within a few milliarcseconds \citep[e.g.][]{Norris93,Mos02} and therefore can be used in combination to reveal the physical conditions in the star formation regions they are detected towards, as the conditions required to produce the two transitions are similar, but not identical \citep{Cragg05}. This means that the presence or absence of accompanying 12.2-GHz emission may be determined by only a small change in physical conditions, which has been proposed to be a reflection of the evolutionary stage of the  associated high-mass star formation region \citep[e.g.][]{Ellingsen07,Breen10a}.

Galactic water masers have been detected towards regions of star formation with masses extending to low-mass objects \citep[e.g.][]{Claussen96,Furuya01} as well as evolved stars \citep[e.g.][]{Deacon07}. Water masers can exhibit extreme levels of temporal variability \citep[e.g.][]{Brand2003,Felli07}, meaning that single-epoch searches can misrepresent true detection statistics. Combined, the tendency for water masers to be associated with a range of exciting objects and their temporal variability have hampered efforts to study a complete population of water masers associated with high-mass stars. 

The relative occurrence of 6.7-GHz methanol masers and water masers has been previously investigated through targeted observations \citep[e.g.][]{Beuther02,Szymczak2005,Xu08,Breen10b,Titmarsh14,Titmarsh16}. Three of these significant investigations \citep{Szymczak2005,Titmarsh14,Titmarsh16} targeted their water maser observations towards statistically complete samples of 6.7-GHz methanol masers, finding water maser detection rates of $\sim$50 per cent. \citet{Xu08} made 6.7-GHz methanol maser observations towards known 22-GHz water masers, yielding a detection rate of $\sim$35 per cent (once low-mass objects were excluded). \citet{Beuther02} targeted their 6.7-GHz methanol and water maser observations towards 29 high-mass star forming regions, finding that $\sim$38 per cent of methanol masers show no water maser emission and $\sim$35 per cent of water masers show no methanol maser emission. \citet{Breen10b} conducted a sensitive Australia Telescope Compact Array (ATCA) search of 270 6.7-GHz methanol masers, finding a water maser detection rate of 73 per cent. In these previous studies comparing 6.7-GHz and water maser occurrence, biases in the way the targets were selected, and indeed the fact that the observations were targeted at all, have heavily affected the detection statistics, resulting in association rates between 35 and 73 per cent. 

Likewise, until the MMB survey the detection rate of 12.2-GHz methanol maser emission towards 6.7-GHz sources varied from 19 to 60 percent \citep{Gay,Caswell95b,Blas04,Breen10b} depending on the sample selection and the sensitivity of the observations. Since the 12.2-GHz search that targeted the MMB 6.7-GHz maser sample, we now understand that the true detection rate is $\sim$45 per cent \citep{Breen16}. Studies of large, complete samples of excited-state OH masers are much rarer, perhaps limited to the MMB itself \citep{Avison16}.

Following the completion of both the Methanol Multibeam (MMB) Survey \citep{Green09} and the H$_2$O Galactic Plane Survey \citep[HOPS;][]{Walsh11} we are now in a position to conduct a definitive comparison between statistically complete sample of the most common masers found towards high-mass star forming regions. Here we present a comparison of the occurrence of 6.7-GHz methanol masers, 12.2-GHz methanol masers, water masers and 6035-MHz excited-OH masers located within a 100 square degree portion of the southern Galactic plane made possible by the MMB and HOPS surveys described in Sections~\ref{sect:mmb} and~\ref{sect:hops}.

\subsection{The Methanol Multbeam (MMB) Survey}\label{sect:mmb}

The MMB survey searched the portion of the Southern Galactic plane visible to the Parkes 64-m radio telescope. While the primary target of the survey was the 6.7-GHz methanol maser line, a concurrent survey for 6035-MHz OH masers was also preformed \citep[see][for a description of the survey or Table~\ref{tab:surveys} for a summary of the survey parameters]{Green09}. In total, 972 methanol masers and 127 excited-state OH masers were detected and are presented in a series of catalogue papers \citep{CasMMB10,GreenMMB10,CasMMB102,Green12,Breen15,Avison16}. Each of the newly discovered 6.7-GHz methanol and excited-state OH masers, together with any known sources that had not been previously observed with an interferometer where subsequently observed with either the ATCA or MERLIN between 2006 and 2014, achieving positional accuracies better than 0.4 arcsec. Each of the detected masers was re-observed with Parkes in a sensitive `MX' observation (where the on-source position is cycled through each of the seven beams of the receiver) that had an rms noise of $\sim$0.07~Jy, and these data are typically what is used in the following analysis. The initial survey and `MX' observations were chiefly conducted between 2006 and 2009. 

All of the MMB methanol masers have been targeted, using the Parkes 64-m radio telescope, for accompanying 12.2-GHz methanol maser emission \citep{BreenMMB12a,BreenMMB12b,BreenMMB14,Breen16}. Since these masers are always found with 6.7-GHz methanol maser counterparts (and are generally weaker) a complete census of the population has been gained through the targeted observations of MMB sources. A total of 438 12.2-GHz methanol masers were detected towards the MMB sources, equating to a detection rate of 45.3 per cent \citep{Breen16}. See Table~\ref{tab:surveys} for a summary of the survey parameters.

\begin{table*}
 \caption{Survey parameters of the surveys described in Sections~\ref{sect:mmb} and~\ref{sect:hops}. Column 1 indicates the survey name, column 2 gives the telescopes used in those surveys (here P = Parkes, A = ATCA, M = MERLIN, Mop = Mopra). Column 3 gives the maser targeted in the survey, column 4 gives the Galactic coverage of the survey, column 5 gives the year of the initial survey observations in the 2000s, column 6 gives the typical 5-$\sigma$ values of each of the surveys, column 7 gives the values at which the MMB and HOPS surveys are estimated to be complete at, column 8 gives the velocity resolution of the survey data, and column 9 gives the years that the data used in the analysis was chiefly collected in (since for all but the 12.2-GHz survey, it is subsequent, more sensitive follow-up data used in this analysis).}
  \begin{tabular}{lcccccccc} \hline
 \multicolumn{1}{c}{\bf Survey} &\multicolumn{1}{c}{\bf Telescope}&\multicolumn{1}{c}{\bf Maser} &\multicolumn{1}{c}{\bf Coverage}  &  {\bf Initial} &\multicolumn{1}{c}{\bf 5-$\sigma$} & {\bf Complete} &{\bf Vel res} & {\bf Epoch} \\
  & & & & {\bf survey}&{\bf (Jy)} &{\bf (Jy)} & {\bf (\kmsns)} & {\bf (yr)} \\
 
 & & & & {\bf (yr)}& & &  &    \\
  \hline

MMB	& P, A, M &	6.7-GHz meth		&	186$^{\circ}$$<$$l$$<$60$^{\circ}$,	$b$=$\pm$2$^{\circ}$ & 06 - 09 &	0.85 & 1& 0.11 & 06 - 09 	\\
	& P, A		& 6035-MHz OH & 186$^{\circ}$$<$$l$$<$60$^{\circ}$,	$b$=$\pm$2$^{\circ}$ & 06 - 09&0.85& &  0.12 &06 - 09  \\
12.2-GHz  & P &12.2-GHz meth 		&  MMB targeted   & 08, 10, 15 & 0.85 & & 0.08  & 08, 10, 15\\
HOPS		& Mop, A	&  22-GHz water 	&	290$^{\circ}$$<$$l$$<$30$^{\circ}$, b$\pm$0.5$^{\circ}$ & 08 - 10& 2 	& 10 & 0.42 &11, 12 \\
\hline
\end{tabular}\label{tab:surveys}
\end{table*}

\subsection{The H$_2$O southern Galactic Plane Survey (HOPS)}\label{sect:hops}

HOPS observed 100 square degrees of the Southern Galactic plane for a number of spectral lines between 19.5 and 27.5 GHz, including water masers \citep{Walsh11}. Observations were made outside the traditional mm season which resulted in a slightly variable detection limit across the survey region. 
All water maser detections were followed up with more sensitive ATCA observations to derive precise positions \citep{Walsh14}. See Table~\ref{tab:surveys} for a summary of the survey parameters.

A total of 2790 water maser spots were detected, distributed amongst 631 maser sites (defined by \citet{Walsh14} to be any spots enclosed within a 4 arcsec radius). A total of 31 masers detected in the initial survey were not detected during the ATCA follow up observations (consistent with expectations of water maser variability), meaning that 122 additional sites were detected in the ATCA observations. These additional sources can easily be accounted for by the vast improvements in both spatial resolution and sensitivity.
 
Of the 631 water masers detected by \citet{Walsh14}, 121 (19 per cent) are associated with evolved stars and are therefore excluded from our sample. \citet{Walsh14} also designated 77 (12 per cent) of their water maser sources as unknown but we found that two of these are associated with MMB sources and have therefore changed their designation to reflect that they are associated with star formation regions. In the following analysis we consider only the 435 water masers found to be associated with star formation regions. The \citet{Walsh14} star formation designations do not discriminate on the basis of mass range but the vast majority of the water masers designated as being associated with star formation here will be associated with high-mass star formation. If HOPS has detected water masers towards low mass stars, they would fall into the unknown category since the star formation designations rely on the association with 6.7-GHz methanol masers or infrared characteristics indicative of high-mass star formation.

\section{Determining associations between the MMB and HOPS}

Determining associations between water and other masers can be difficult, due to the intrinsic difference in spot distributions, site definitions and positional uncertainties. The spots associated with a methanol maser site tend to be located within a fairly compact region, typically contained within less than 2 arcsec \citep{C09,CasMMB10} while water maser emission can be spread over 4 arcsec (corresponding to corresponding to 0.1~pc at a distance of 5~kpc) or more \citep[e.g.][]{R88}. The tendency for water maser emission to be more scattered is due to their collisional excitation, allowing spots to be associated with outflows, for example, and causing them to be spread over greater extents. This presents a challenge when assigning individual spots to a water maser site, especially given the large populations of water masers and crowded star formation regions. To combat this, HOPS reported individual maser spots as well as categorising spots into sites, which were defined by an upper limit on the spot distribution of 4 arcsec. The positional uncertainties are generally much less of a concern: the uncertainty of the MMB survey is 0.4 arcsec \citep[e.g.][]{CasMMB10} and the water maser positions from HOPS are expected to be accurate to around one arcsec.

\begin{figure}\vspace{-1.5cm}
	\epsfig{figure=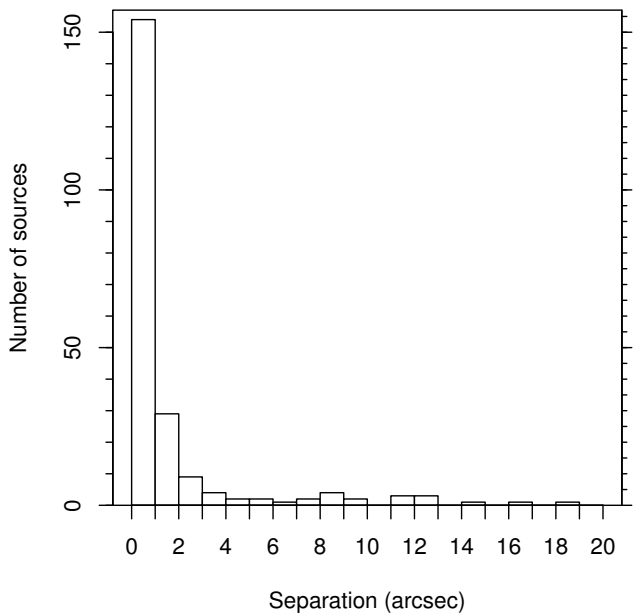,height=9cm,angle=270}
\caption{Separation between MMB source and nearest water maser spot reported in HOPS within a 20 arcsec threshold. }
\label{fig:sep}
\end{figure}

To account for the differences in spot distributions and site definitions, we have compared the MMB methanol maser site positions with the HOPS spot positions in Fig.~\ref{fig:sep}. This comparison should allow us to capture more true associations as the separations will not be dominated by large water maser spreads {\color{magenta} since we are considering every spot individually}. There are 218 MMB sources that have a HOPS water maser spot within 20 arcsec. Fig.~\ref{fig:sep} shows that the majority of these (183) are separated by $\leq$2 arcsec. The remaining 35 MMB sources are separated by more than 2 arcsec from their nearest HOPS {\color{magenta} water maser spot} and nine of those fall in the 2 - 3 arcsec bin. Since we are comparing the MMB sources with the locations of the individual water maser spots, we have essentially already accounted for the tendency for water masers to be more spread out than their methanol maser counterparts and we therefore set the association threshold at 2 arcsec. While it is possible that some of the nine sources that fall in the 2 - 3 arcsec bin are truly associated (in e.g. the case where the only water maser spots detected are those associated with either the red- or blue-shifted component of an outflow), it is most likely that these masers are associated with different individual objects in a crowded star forming region. 

We apply the same method and threshold to determine associations between water masers and excited-state OH masers. 6.7-GHz (and 12.2-GHz) methanol masers and excited-state OH masers are considered to be associated if their site positions are within 2 arcsecs of one another. 

\section{Population statistics}

Within the HOPS survey range we can compare the detection statistics of complete surveys for 6.7-GHz and 12.2-GHz methanol masers, 6035-MHz excited-state OH masers and water masers. Table~\ref{tab:stats} shows that within the 100 square degrees, there are 634 6.7-GHz methanol masers, 435 22-GHz water masers associated with star formation, 295 12.2-GHz methanol masers and 80 excited-state OH masers. As discussed in Section~\ref{sect:limits}, the lower sensitivity of the HOPS survey results in an underestimation of the number of water masers, and if that survey had a detection limit comparable to the other searches, it is likely that it would have found hundreds more sources in the survey range (perhaps exceeding the number of 6.7-GHz sources).

\begin{figure}
{\includegraphics[scale=0.5]{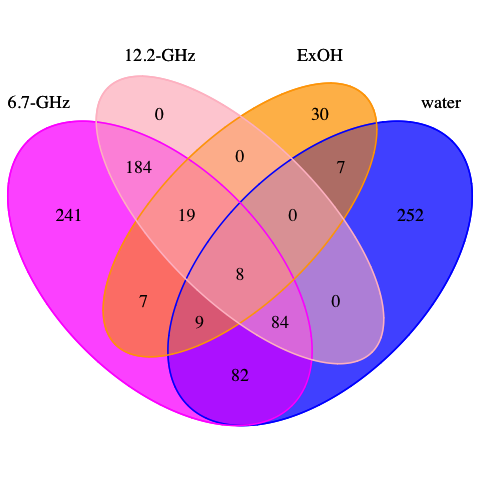}}
{\caption{Venn diagram showing the associations between the four maser species: 6.7-GHz and 12.2-GHz methanol, excited-state OH and water. Note that five water masers have spots are associated with two 6.7-GHz methanol masers, and one water maser has spots associated with three 6.7-GHz methanol masers (and some of these are also associated with 12.2-GHz methanol or excited-state OH emission) and therefore the number of water masers shown in this diagram is 442 (i.e. seven more than the total number of water masers).}\label{fig:venn}}
 \end{figure}

Table~\ref{tab:stats} gives the number of masers of each transition detected in the 100 square degree region and Fig.~\ref{fig:venn} shows the associations between the species. We find that 259 of the star formation water masers are devoid of methanol maser counterparts, meaning that 40 per cent of star formation water masers have a methanol maser counterpart. The complementary statistic is that 28 per cent of the 6.7-GHz methanol masers have associated water masers detected in HOPS. 

 The percentage of 6.7-GHz methanol masers that have associated water maser emission is approximately the same whether or not the 6.7-GHz emission is accompanied by 12.2-GHz emission (31 per cent) or not (27 per cent). \citet{BreenMMB14} compared the occurrence of water masers and 12.2-GHz methanol masers in the 6$^{\circ}$ to 20$^{\circ}$ longitude range, finding that water masers were more often found towards 6.7-GHz methanol masers with accompanying 12.2-GHz methanol maser emission (detection rate of 55.7 percent compared to 36.2 percent for 6.7-GHz methanol masers devoid of 12.2-GHz emission). The main difference between the samples is a vastly different water maser detection limit, suggesting that the less sensitive HOPS sample is missing a higher relative number of water masers that are associated with both 6.7-GHz and 12.2-GHz sources. In their observations of water masers towards 6.7-GHz methanol masers, \citet{Titmarsh16} detected a smaller proportion of water masers with flux densities less than 2~Jy towards 6.7-GHz methanol masers devoid of accompanying 12.2-GHz maser emission (34 per cent) compared to those sources with both 6.7- and 12.2-GHz emission (47 per cent). Although the sample size is inadequate to draw firm conclusions (34 water masers accompanying both 6.7- and 12.2-GHz emission and 67 associated with just 6.7-GHz methanol maser emission) this is consistent with relatively more water masers associated with both 6.7- and 12.2-GHz falling below the detection limit of the HOPS observations.

Just over half (54 per cent) of the 80 excited-state OH masers are associated with 6.7-GHz methanol masers, and are more commonly associated with 6.7-GHz methanol masers that also have 12.2-GHz emission (27 compared to 16) or sources that have any combination of 2 or more other masers (36 compared to 14). This is consistent with the notion that OH masers are associated with a slightly later phase of the star formation process \citep[e.g.][]{C97}.

\begin{table}\footnotesize
 \caption{Detection statistics of the different maser species within the HOPS survey range.}
  \begin{tabular}{lccc} \hline
 \multicolumn{1}{c}{Maser type} &\multicolumn{1}{c}{\# of} &\multicolumn{1}{c}{\# of solitary}  & \multicolumn{1}{c}{percentage} \\
 &\multicolumn{1}{c}{detections} &\multicolumn{1}{c}{sources} &\multicolumn{1}{c}{solitary} \\ \hline

6.7-GHz methanol 	&	634		&	241	&	38	\\
SF water masers	&	435		&	252	&	58	\\
12.2-GHz methanol	&	295		&	0	&	0	\\
Excited-OH masers	&	80		&	30	&	38	\\
\hline
\end{tabular}\label{tab:stats}
\end{table}

\subsection{The effect of different detection limits and source variability}\label{sect:limits}

While the maser surveys we are comparing are all complete over the same range of Galactic longitude, the HOPS survey has a much higher detection limit than the others which means that the number of water masers (and therefore water maser associations) should be considered a lower limit. Even with the higher detection limit, HOPS is still a valuable resource, being the largest systematic search for water maser emission ever conducted. It is also worth noting that even though the searches for excited-state OH, 6.7-GHz and 12.2-GHz methanol masers have similar detection limits, the luminosity functions of these different types of masers are different \citep[well known in the case of the 6.7- and 12.2-GHz sources, e.g.][]{Caswell95b,Breen12stats} and as such the comparison between the populations will be imperfect. 

A dedicated water maser follow-up to the MMB survey conducted with the ATCA \citep{Titmarsh14,Titmarsh16}, found water maser emission towards 48 per cent of the 323 MMB masers in the longitude range 341$^{\circ}$ (through the GC) to 20$^{\circ}$. The detection limit of this survey was $\sim$40 times lower than the completeness level of the HOPS survey and has therefore detected a higher fraction of the water maser population towards 6.7-GHz methanol masers. Our detection statistics show that 28 per cent of 6.7-GHz methanol masers have an associated HOPS water maser. This implies that HOPS was able to recover about 58 per cent of the water masers towards a sample of 6.7-GHz methanol masers that a more sensitive survey would. It is difficult to estimate how many solitary water masers were missed by HOPS since there is no large scale, sensitive survey available for comparison. If we assume that the luminosity distribution of solitary water masers is similar to the distribution of water masers associated with 6.7-GHz methanol masers then we would also expect that $\sim$42 per cent of sources are missing. However, previous studies have shown that solitary water masers tend to have lower peak flux densities than those associated with 6.7-GHz methanol masers \citep{Breen10b,BE11}, so the sensitivity limitations of HOPS will have prevented an even higher percentage of these solitary water masers from being detected. 

A further difficulty presented by water maser detection statistics is that a single epoch search will not recover the entire population of sources due to their high level of temporal variability. HOPS itself is a good example of this; the initial survey with the Mopra radio telescope detected 540 water maser sites \citep{Walsh11} and 31 of those sources were not detected in the much more sensitive ATCA follow-up observations. This indicates that $\sim$6 percent of the population detected in the initial survey fell below the detection limit of more sensitive observations conducted up to four years later. This percentage implies that, in the case of HOPS, the sensitivity of the initial survey has more effect on the detected population than source variability on the timescale of a few years. This is largely a consequence of the high detection limit of HOPS and the more sensitive follow-up observations. It is possible that, again, solitary water masers are more effected by variability than those associated with other types of masers. \citet{Breen10b} found, in a two epoch search, that more solitary sources were detectable only at a single epoch compared to water masers that are associated with either 6.7-GHz or OH maser emission. This means that there is potential for a single epoch search to miss relatively more solitary water masers compared to water masers with methanol or OH counterparts. 

Although at a lower level, variability can also effect the detection of methanol and excited-state OH masers which will also have some effect on the comparison of these maser populations. We can be relatively confident that few 6.7- and 12.2-GHz methanol masers are missing from our sample due to source variability. The MMB survey found that less than 0.5 per cent of the 6.7-GHz methanol masers detected, varied enough to fall below the detection limit of follow-up interferometric observations used to pinpoint their locations. Furthermore, to minimise the effects of variability on the population of 12.2-GHz methanol masers, those observations routinely included a second (and sometimes third) observation epoch of non-detections which led \citet{BreenMMB12a} to conclude, from completeness tests, that few non-detections could have emission below the detection limit of the search.

 The excited-state OH maser survey \citep{Avison16} confidently detected 80 sources, but a further 12 known excited-state OH masers were either tentatively detected (6) or went completely undetected (6), indicating that at any one epoch we can estimate that $\sim$7 per cent of the excited state OH maser population have varied below the detection limit of a search of similar sensitivity.

\section{Basic properties of the different maser populations} 

\subsection{Distribution of source velocities}

\begin{figure}\vspace{-1.5cm}
\begin{center}
	\epsfig{figure=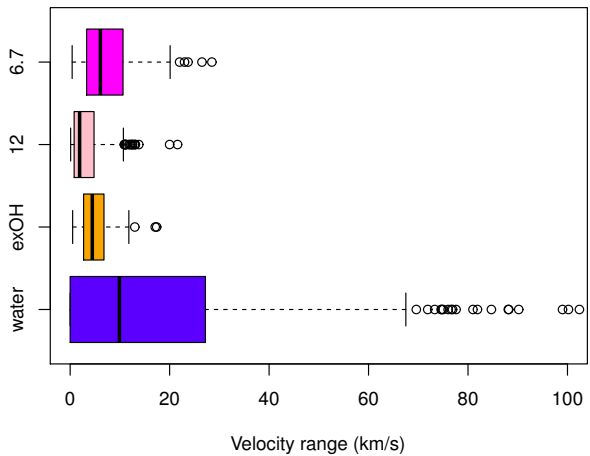,height=9cm,angle=270}
\caption{Velocity range box plots of each of the four maser types; 6.7-GHz methanol (magenta), 12.2-GHz methanol (pink), excited-state OH (orange) and water (blue). Water masers with velocity ranges in excess of 105~\kms (26; with ranges up to 299.6~\kmsns) have been excluded from the plot in order to see the other, smaller ranges more clearly.
In each box plot, the solid vertical black line represents the median of the data, the coloured box
represents the interquartile range (25th to the 75th percentile), and the dashed horizontal lines (the `whiskers') show the range from the 25th percentile to the minimum value and the 75th percentile to the maximum value, respectively. Values that fall more the 1.5 times the interquartile range from either the 25th or 75th percentile are considered to be outliers are represented by open circles.}
\label{fig:vel_box}
\end{center}
\end{figure}

Fig.~\ref{fig:vel_box} shows box plots of the velocity ranges of each of the 6.7-GHz methanol, 12.2-GHz methanol, excited-state OH and water masers. The excited-state OH maser data presented in \citet{Avison16} gives values for LHCP (left hand circularly polarised) and RHCP (right hand circularly polarised) emission separately and we have used the larger of the LHCP or RHCP velocity range in this analysis. The median velocity range is highest for the water masers at 9.8~\kmsns, followed by the 6.7-GHz methanol (6.1~\kmsns), excited-state OH (4.5~\kmsns), and 12.2-GHz methanol masers (1.9~\kmsns). Across the full MMB range, the median velocity of 6.7-GHz methanol masers is 6~\kms \citep{Green17} and the median of 12.2-GHz sources is 1.7~\kms \citep{Breen16} and therefore are consistent with the sub-sample we consider here. The fact that the water masers show the largest velocity range is expected, particularly because of their tendency to trace high-velocity outflow, but the median velocity of the HOPS sources is significantly lower than either of the \citet{Breen10b} or \citet{Titmarsh14,Titmarsh16} targeted water maser observations which have medians of 15 and 17~\kmsns, respectively. A part of this difference can be accounted for by the fact that the \citet{Walsh14} quoted peak velocities of spots and therefore results in an underestimation of the velocity ranges of sites, but some of the difference is due to the unbiased nature of HOPS. The fact that the velocity range of the excited-state OH masers exceed the 12.2-GHz methanol masers is perhaps surprising given that 12.2-GHz methanol masers generally have much higher peak flux densities, and suggests that the excited-OH emission is arising from a larger volume of gas than the 12.2-GHz maser emission.

\citet{Breen16} found that there was a preponderance of 12.2-GHz methanol masers with high velocity ranges within the Galactic longitude ranges of G\,330 - G\,340 and to a slightly lesser extent, G10 - G30. These Galactic longitude ranges correspond to significant structures in the Galaxy and it was suggested that the higher velocity ranges reflected the enhanced star formation resulting in larger relative motions that were then reflected in the individual star forming regions. We have repeated this analysis using the HOPS water masers, and found a similar peak in the G330 - G340 longitude range, but in the case of the water masers, this peak is entirely consistent with the higher number of sources in this longitude range (i.e. $\sim$17 per cent of the water maser population is located in this longitude range, and $\sim$19 per cent of the population of water masers with velocity ranges in excess of the median water velocity range lie in this longitude range). The fact that the velocity range of the water masers do not show a significant overabundance of high velocity sources (as is the case for 12.2-GHz methanol masers) is likely due to the fact that water maser velocities are not as closely tied to the star formation region, instead often depending on the interaction of the young star formation region with the surrounding environment. Future comparisons with dense gas line-widths of star formation regions as a function of Galactic longitude will be able to shed light on this peculiar result. 

\begin{figure*}\vspace{-1cm}
\begin{center}
	\epsfig{figure=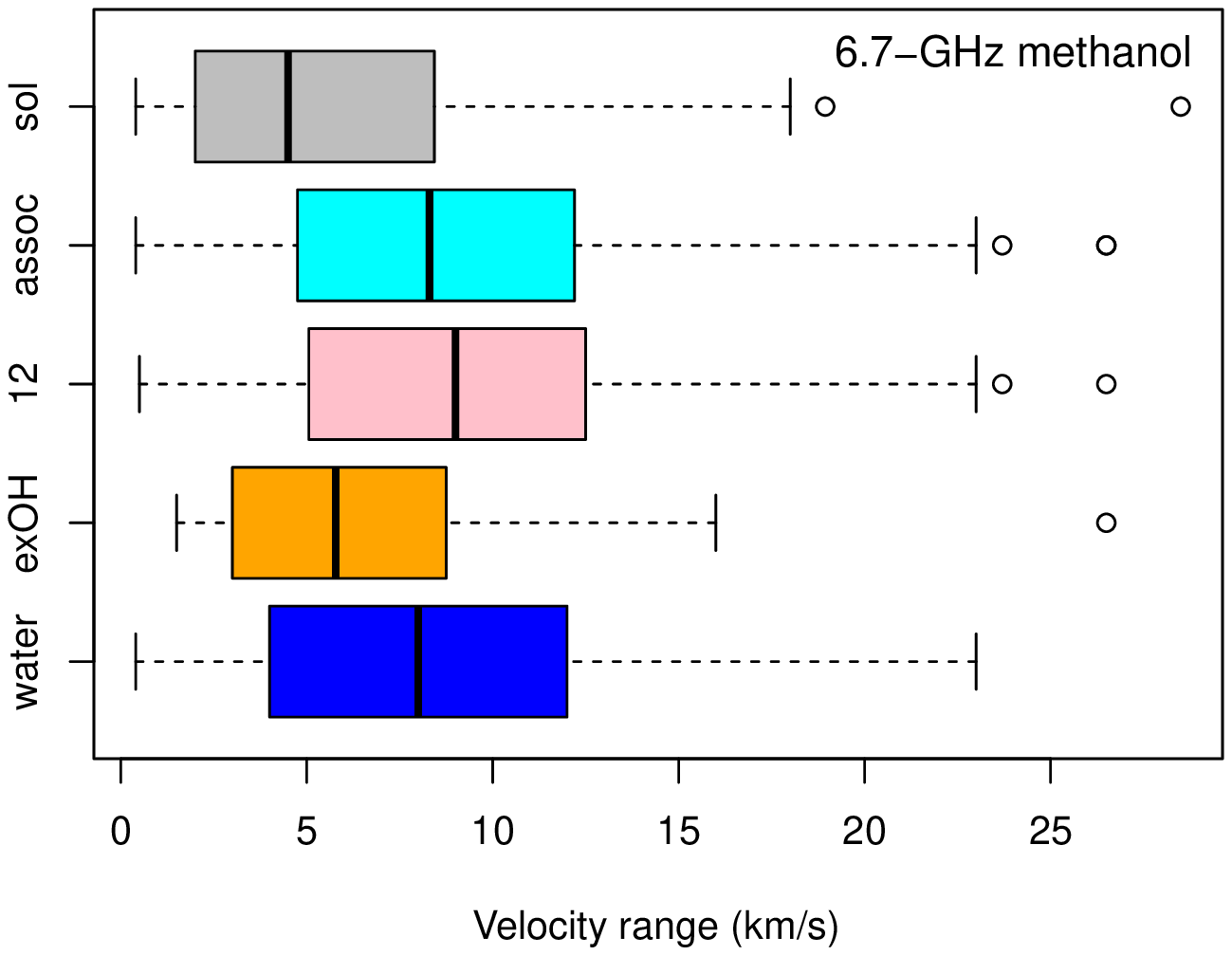,height=8cm,angle=270}\vspace{-1cm}
	\epsfig{figure=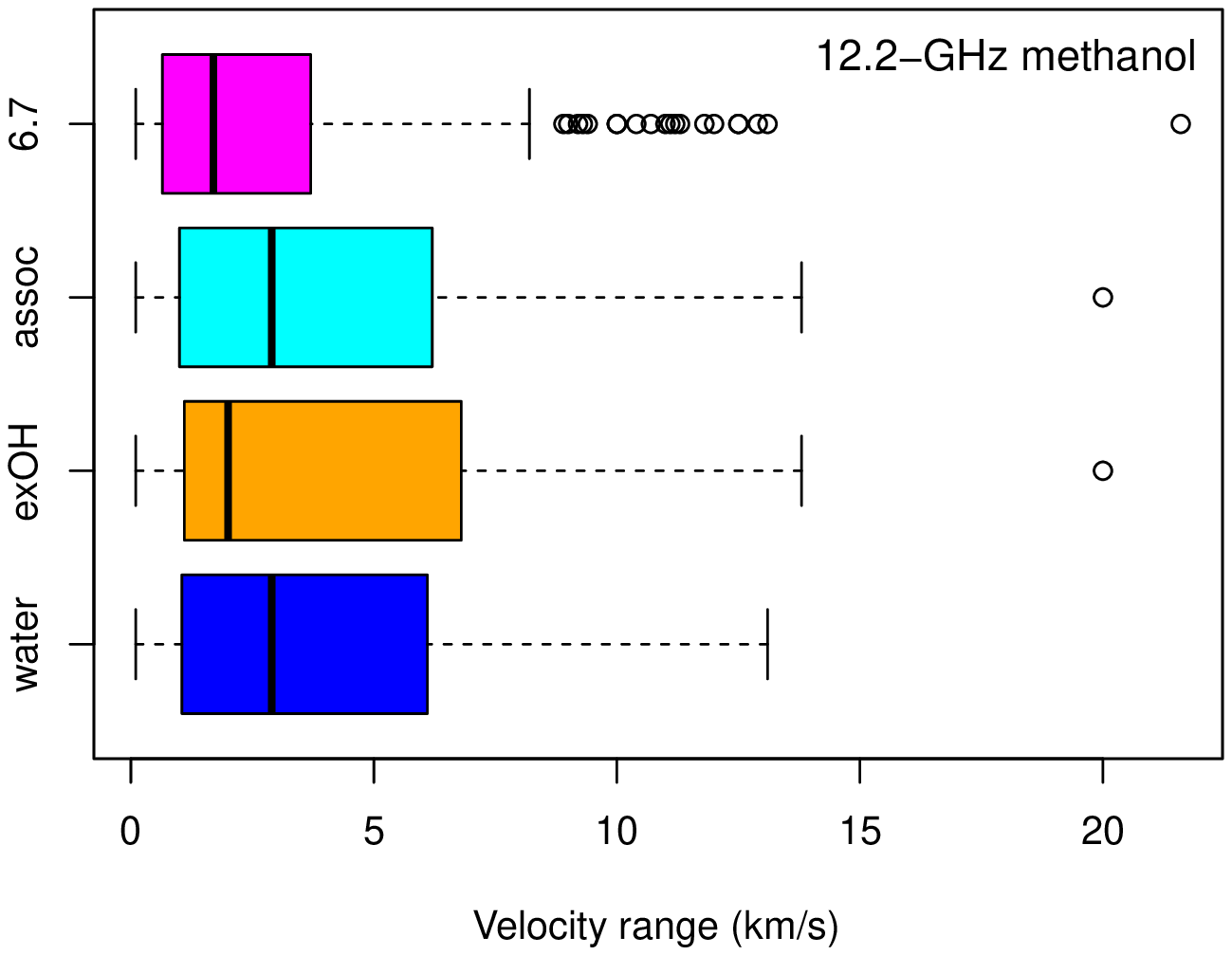,height=8cm,angle=270}
	\epsfig{figure=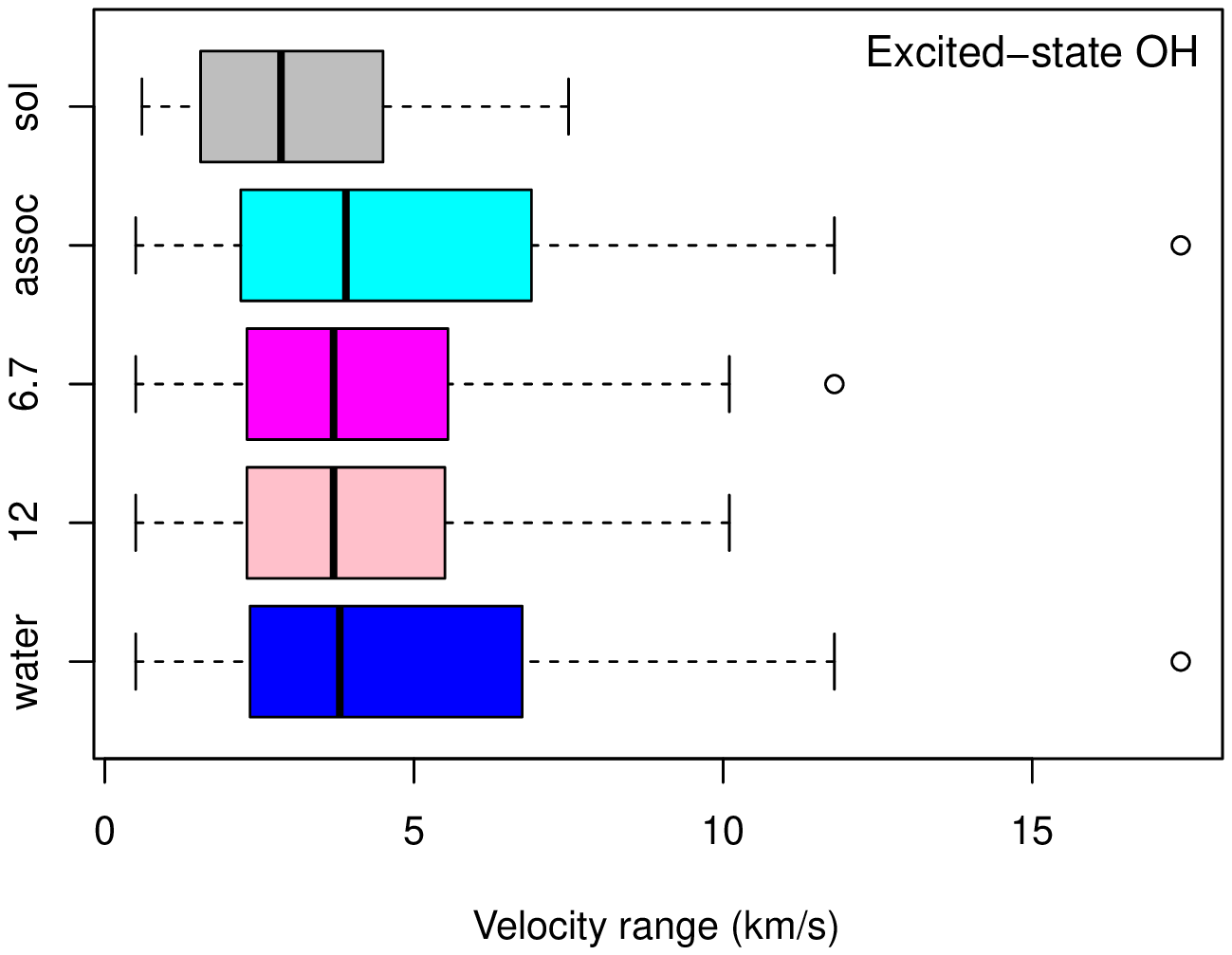,height=8cm,angle=270}
\epsfig{figure=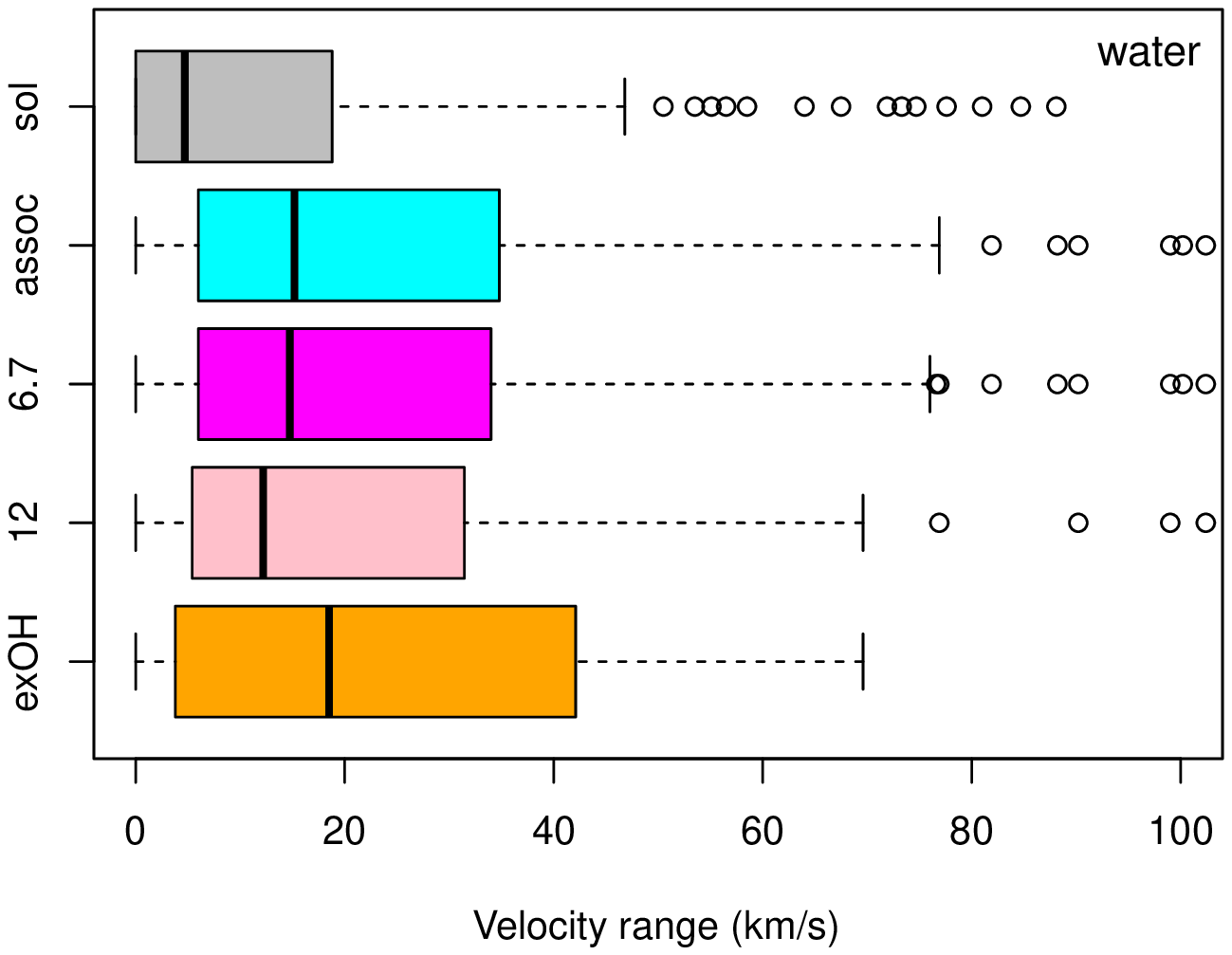,height=8cm,angle=270}
\caption{Velocity range box plots of 6.7-GHz methanol masers (top left), 12.2GHz methanol masers (top right), excited-state OH masers (bottom left) and water masers (bottom right) in the categories of solitary (`sol'; grey), associated with one or more of the other transitions (`assoc'; cyan), associated with 6.7-GHz methanol (`6.7'; magenta), 12.2-GHz (`12'; pink), excited-state OH (`exOH'; orange) and water masers (`water'; blue). Note that in the case of 12.2-GHz methanol masers `assoc' includes those 12.2-GHz sources that are associated with either or both of excited-state OH or water maser emission since all are associated with 6.7-GHz methanol maser sources. The water maser plot has been truncated in order to see detail in the other distributions and the full range of water maser outliers extends to 299.6~\kms (i.e. 11 water maser sources have been excluded from the plot). See Fig.~\ref{fig:vel_box} caption for a general explanation of box plots.}
\label{fig:vel_box_associations}
\end{center}
\end{figure*}

Fig.~\ref{fig:vel_box_associations} shows the velocity ranges of each of the four types of masers split into their association categories of solitary, associated (with any combination of the other three maser types), and associated with each of 6.7-GHz methanol, 12.2-GHz methanol, excited-state OH and water maser emission. In the case of 12.2-GHz methanol masers there are no `solitary' sources since all have 6.7-GHz counterparts so for these sources `associated' means that they have accompanying emission from either or both of excited-state OH and water maser emission. Immediately evident in Fig.~\ref{fig:vel_box_associations} is the fact that `solitary' masers have velocity range distributions that are skewed towards smaller values. This has been noted previously in the case of water masers \citep[e.g.][]{BE11} and for 6.7-GHz methanol masers \citep[e.g.][]{Breen12stats}.

Since \citet{Walsh14} published the peak velocity of each detected maser feature, masers consisting of a single spectral feature have a calculated velocity range of zero. It can be seen in Fig.~\ref{fig:vel_box_associations} that sources with velocity ranges of zero \kms account for a large portion of the `solitary' water masers - in fact, there are 89 (equating to more than a third of sources in that category), compared to 26 ($\sim$14 per cent) in the associated category. \citet{Breen10b} found that water masers with single features, and those that have no 6.7-GHz methanol or OH maser counterpart, are more likely to fall below the detection limit of a sensitive, single-epoch search and so it is likely that this population is particularly underrepresented in the HOPS sample.

\begin{figure}\vspace{-1.5cm}
\begin{center}
	\epsfig{figure=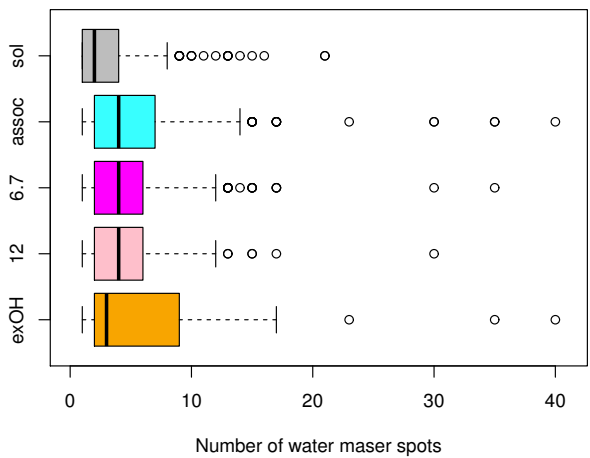,height=9cm,angle=270}
	
\caption{Box plots of the number of water maser spots associated with water maser sites in the categories of solitary (`sol'; grey), associated (`assoc'; blue), associated with: 6.7-GHz methanol masers (6.7; magenta), 12.2-GHz methanol masers (12; pink) and excited-state OH (exOH; orange). Here `solitary' means those water masers that are not associated with any of the other three maser types and `associated' means those water masers that are associated with at least one of the other three maser types. The plot range excludes one source that is associated with a methanol maser (G000.677-0.028, with 61 spots). See Fig.~\ref{fig:vel_box} caption for a general explanation of box plots.}
\label{fig:water_box}
\end{center}
\end{figure}

\citet{Walsh14} presented their water masers as a series of water maser spots. One might expect that the number of water maser spots is closely linked to the velocity range of the maser emission and a comparison of Fig.~\ref{fig:water_box}, which shows the number of water maser spots as a function of associations, with the water maser velocity plot shown in the bottom right of Fig.~\ref{fig:vel_box_associations}, reveals that our data are broadly consistent with that assertion. In particular, it can be seen that the water masers that are associated with excited-state OH masers have the largest velocity ranges and also tend to have higher numbers of spots, while solitary water masers have the lowest velocity ranges and the smallest number of maser spots.

\subsubsection{Comparing methanol maser velocities to water and excited-state OH masers}

Methanol masers at 6.7-GHz are considered good probes of systemic velocities, with their central velocities most often falling within $\pm$3~\kms of the systemic velocities of the regions they are associated with \citep[e.g.][]{Szy07,C09,Pandian09,GM11}. Water masers, on the other hand, are generally considered to be poor tracers of systemic velocities due to their association with outflows. We have compared the velocity of the peak water maser emission to the central 6.7-GHz methanol maser emission and find that $\sim$89 per cent of water masers show their peak emission within $\pm$10~\kms of the central 6.7-GHz methanol maser velocity. Of the 11 per cent of sources with velocities at larger separations, we find that 15 sources are more blue-shifted and 6 sources are more red-shifted. Similarly, in the water maser observations targeting MMB sources, \citet{Titmarsh16} found that 88 per cent of water masers showed their peak emission within $\pm$10~\kms of the methanol maser peak, while in observations targeting large, but incomplete samples of 6.7-GHz methanol and OH masers, \citet{Breen10b} found that 78 per cent of their sources shared peak velocities within $\pm$10~\kms of each other. The consistency with the \citet{Titmarsh16} sample is probably reflective of the fact that they also use a portion of the complete sample of MMB masers, but it is interesting that the percentage of sources remains constant with the much more sensitive \citet{Titmarsh16} sample, suggesting that the distribution holds to lower water maser peak flux densities. 

Considering that the central velocity (the middle of the maximum and minimum velocity) of 6.7-GHz methanol masers is a more accurate tracer of systemic velocities than the velocity of the peak emission \citep[e.g.][]{GM11}, we repeated the comparison using water maser central velocity. In this case $\sim$74 per cent of the water maser central velocities fell within $\pm$10~\kms of the central 6.7-GHz velocity and of the 47 sources with greater separations, 29 were more blue-shifted and 18 were more red-shifted. So we conclude that the water maser central velocity is a poorer representation of systemic velocities than the water maser peak velocity.

\citet{CP08} and \citet{CB10} suggested that water masers dominated by blue-shifted emission might represent a short-lived evolutionary phase in the star formation process. To investigate this we have looked at the number of water maser sources with peak velocities offset by more than 10~\kms from the associated 6.7-GHz methanol maser central velocity in the association categories of with and without accompanying 12.2-GHz emission (they all have to have 6.7-GHz emission so we have a measure of the systemic velocity) indicating less and more evolved sources, respectively \citep[e.g][]{Breen10a}. We find that in the case of the sources with associated 12.2-GHz sources, there are six sources blue-shifted by more than 10~\kms and four that are red-shifted by 10~\kmsns, compared with nine and two in the case of sources with no accompanying 12.2-GHz emission, lending support to the suggestion of \citet{CP08} and \citet{CB10} that water masers showing dominant blue-shifted emission are indicative of a particularly young high-mass star formation region (although we note we are only dealing with a small number of sources).

In general, the average of the LHCP and RHCP peak velocities of excited-state OH masers are in much closer agreement with the central velocity of 6.7-GHz methanol masers. Of the 43 excited state OH masers with associated 6.7-GHz masers, 86 per cent of the sources have peak velocities within $\pm$5\kms \citep[][found similar results using their full sample]{Avison16} and only one source, G\,336.983$-$0.183, has a separation of more than 10~\kms (at 17.1~\kms). It is possible that this indicates that the 6.7-GHz methanol maser emission is only aligned with the excited-state OH maser by chance even though their measured angular separations is only 0.4$\arcsec$. Among the other five sources with velocity separations of 5~\kms or more, only two have angular separations greater than 0.4$\arcsec$, at 0.9 and 1.9$\arcsec$, respectively. All six of the excited-state OH sources that have peak velocities separated by more than $\pm$5~\kms from the central 6.7-GHz velocity are redshifted.

\subsection{Distribution of peak and integrated flux densities}

Within the 100 square degree region of the Galactic plane, the only two masers with peak flux densities greater than 1000~Jy are the 6.7-GHz methanol masers G9.621+0.196 and G323.740$-$0.263 with peak flux densities of $\sim$5200~Jy and $\sim$3200~Jy, respectively. There are 26 6.7-GHz methanol masers which had peak flux densities in excess of 100~Jy in their final MX observations (32 in the survey cube; all of which had peak flux densities greater than 82 Jy in the final MX observation). Similarly there are 30 water masers with peak flux densities greater than 100~Jy (9 solitary and 21 associated with at least one other maser species) and G\,12.680-0.182 has the highest peak flux density at 580~Jy. There are six 12.2-GHz methanol masers with flux densities in excess of 100~Jy and the largest peak flux densities are associated with G\,9.621+0.196 and G\,323.740$-$0.263 (also the highest flux density 6.7-GHz sources) with peak flux densities of 401 and 396~Jy, respectively. None of the excited-state OH masers have peak flux densities of more than 100~Jy, the strongest is G\,323.459$-$0.079 (associated with both 6.7- and 12.2-GHz methanol maser emission) at 44~Jy. Only 8 ex-OH sources measured more than 10 Jy in either RHCP or LHCP.

\begin{figure}\vspace{-1.5cm}
\begin{center}
	\epsfig{figure=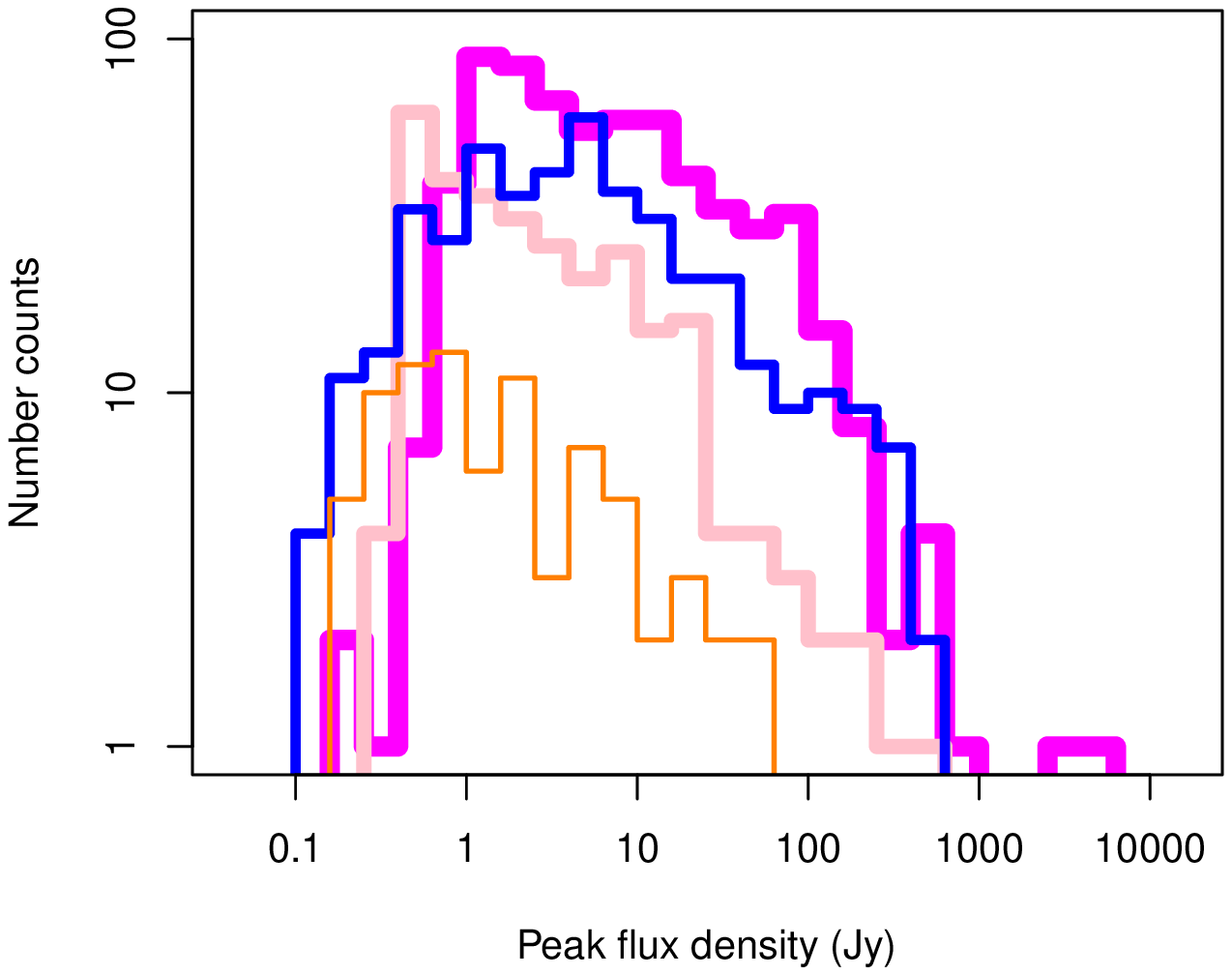,height=9cm,angle=270}\vspace{-1cm}
	\epsfig{figure=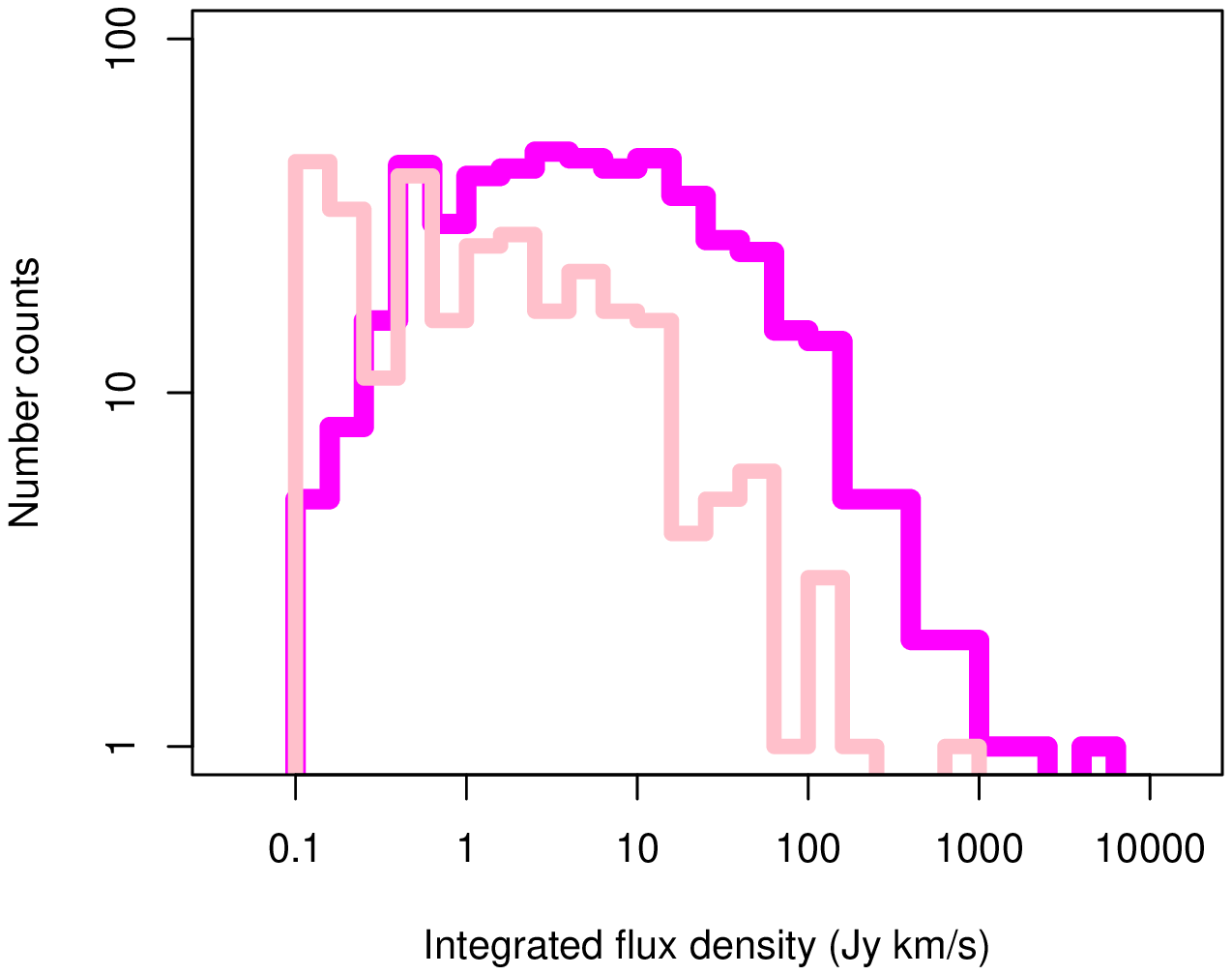,height=9cm,angle=270}
\caption{(Top) Peak flux density distribution of the 6.7-GHz methanol \citep[magenta;][]{CasMMB10,GreenMMB10,CasMMB102,Green12,Breen15}, 12.2-GHz methanol \citep[pink;][]{BreenMMB12a,BreenMMB12b,BreenMMB14,Breen16}, water \citep[blue;][]{Walsh14} and excited-state OH masers \citep[orange;][]{Avison16}, and;  
(bottom) integrated flux density distribution of the 6.7-GHz methanol \citep[magenta;][]{Breen15} and 12.2-GHz methanol \citep[pink;][]{BreenMMB12a,BreenMMB12b,BreenMMB14,Breen16}. }
\label{fig:flux_dist}
\end{center}
\end{figure}

Fig.~\ref{fig:flux_dist} shows the distributions of each of the four maser populations as a function of peak flux density and integrated flux density (in the case of the 6.7- and 12.2-GHz methanol masers). The distribution of the full 6.7-GHz methanol maser population detected in the MMB survey are shown in figs 1 and 2 of \citet{Green17}. Each of the maser types show a turnover at low peak flux densities between $\sim$0.5 and a few Jy. \citet{Green17} investigated the turnover in the 6.7-GHz methanol maser counts and found that after taking incompleteness and statistical errors into account that the turnover was not significant. We expect that this is true for the other three maser types given that the turnovers all occur below the 5-$\sigma$ noise levels of each of the searches. In the case of the 6.7-GHz methanol masers, the possibility of a real flux density turn over will be addressed by the more sensitive `piggyback' survey \citep{Ellingsen17} which was described in \citet{Green09}. Given the sensitivity limitations it is difficult to confidently comment on any differences in the peak flux density distributions beyond the obvious ranking in peak flux densities from 6.7-GHz methanol, water, 12.2-GHz methanol and, finally, excited-state OH. 

The lower panel of Fig.~\ref{fig:flux_dist} shows the distributions of integrated flux densities for the 6.7- and 12.2-GHz methanol masers (these values are not available for either the water masers or the excited-state OH masers) and the overall distributions are similar to that of the peak flux densities. The most striking difference in the distributions is the preponderance of 12.2-GHz methanol masers with low integrated flux densities, dominated by a large number of weak, single feature masers which is not replicated in the 6.7-GHz distribution.

\begin{figure*}\vspace{-1.5cm}
\begin{center}
	\epsfig{figure=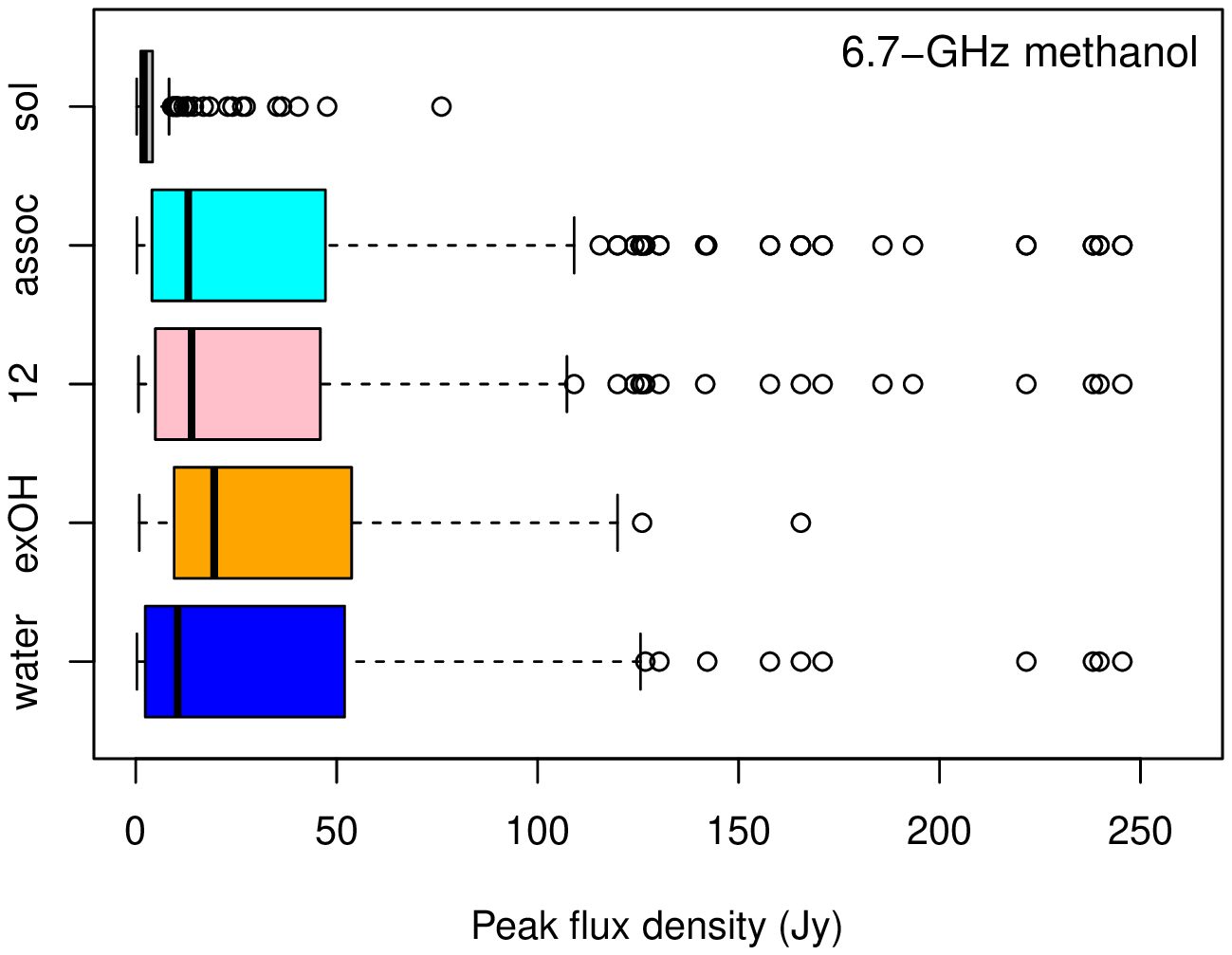,height=8cm,angle=270}\vspace{-1cm}
	\epsfig{figure=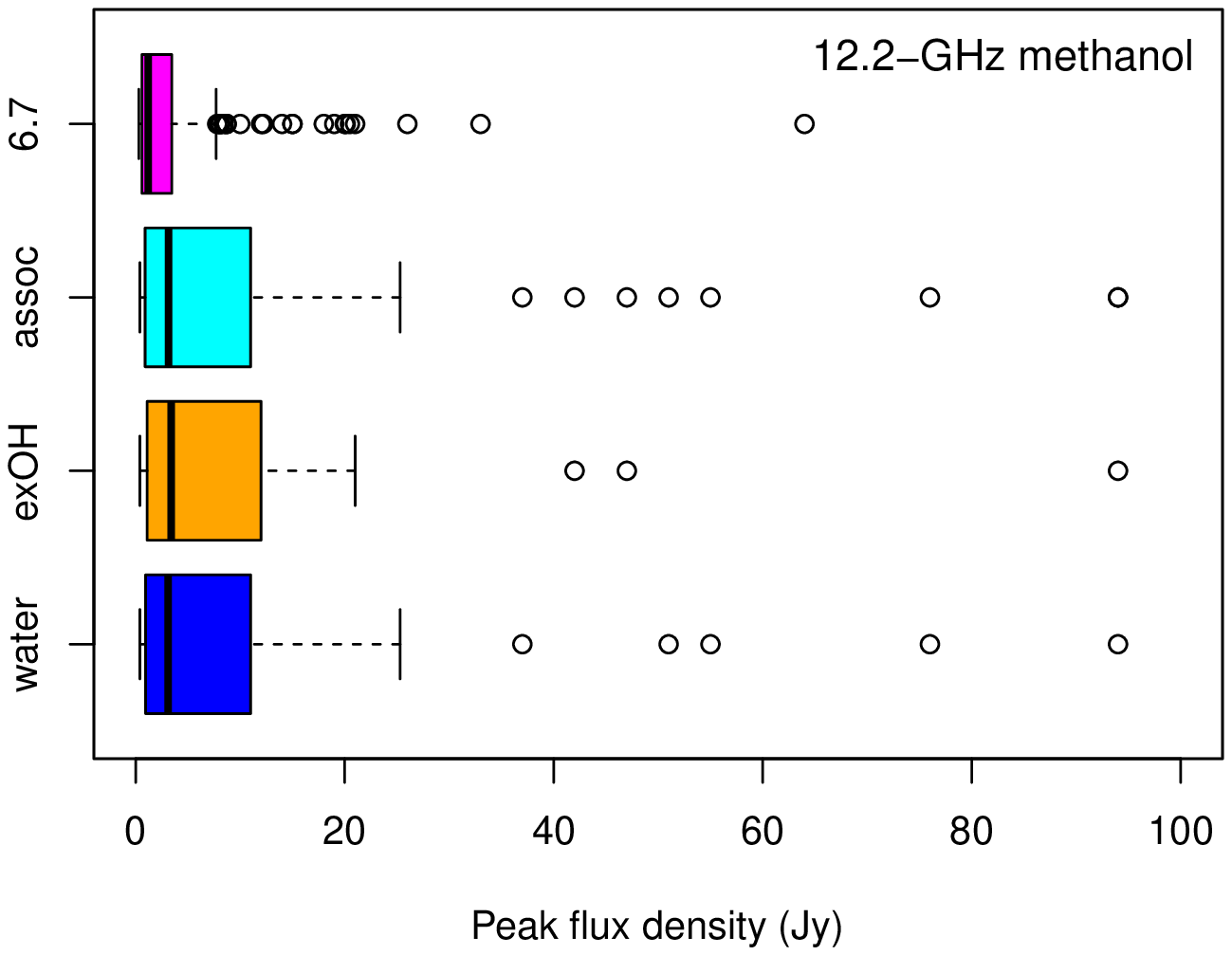,height=8cm,angle=270}
	\epsfig{figure=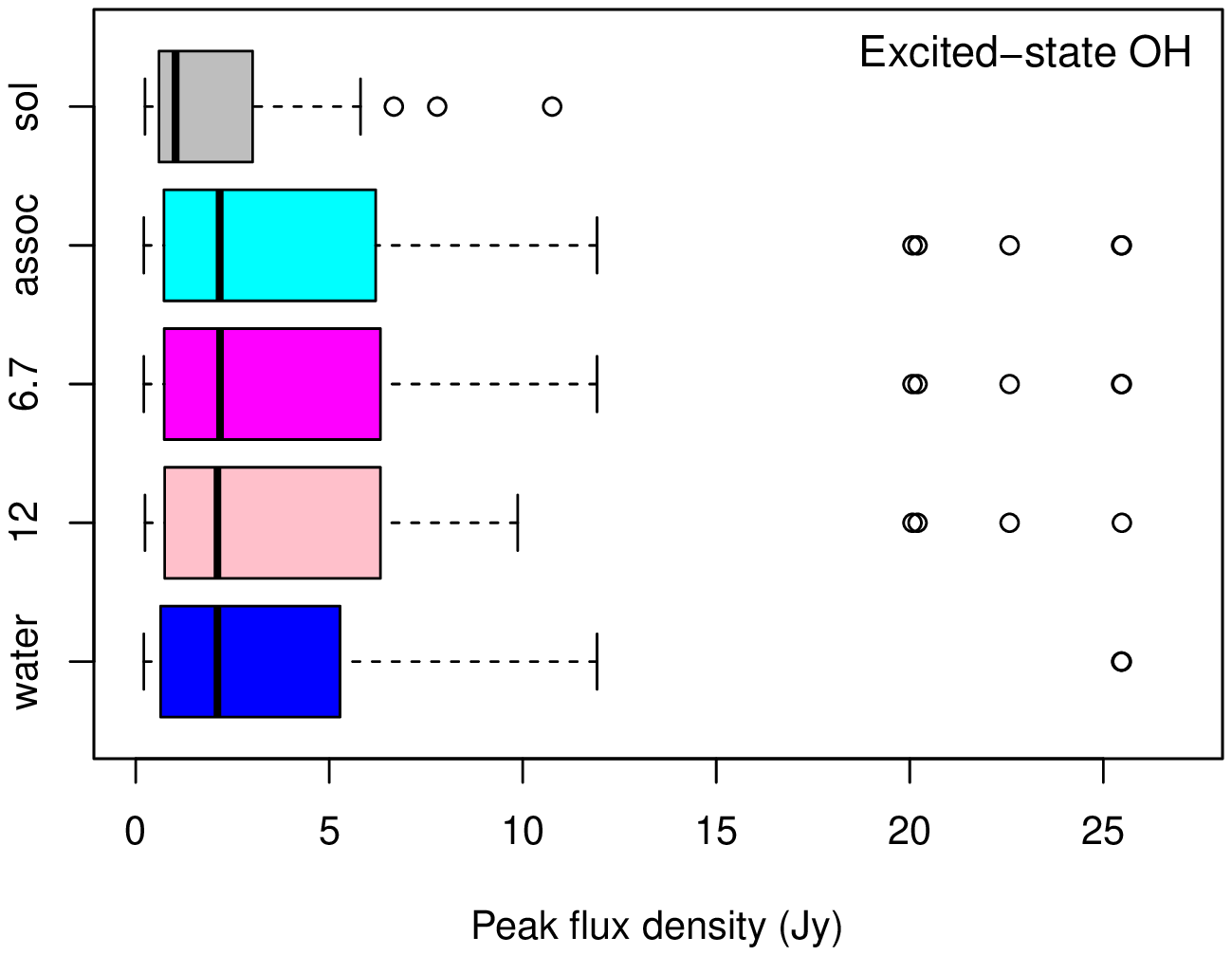,height=8cm,angle=270}
\epsfig{figure=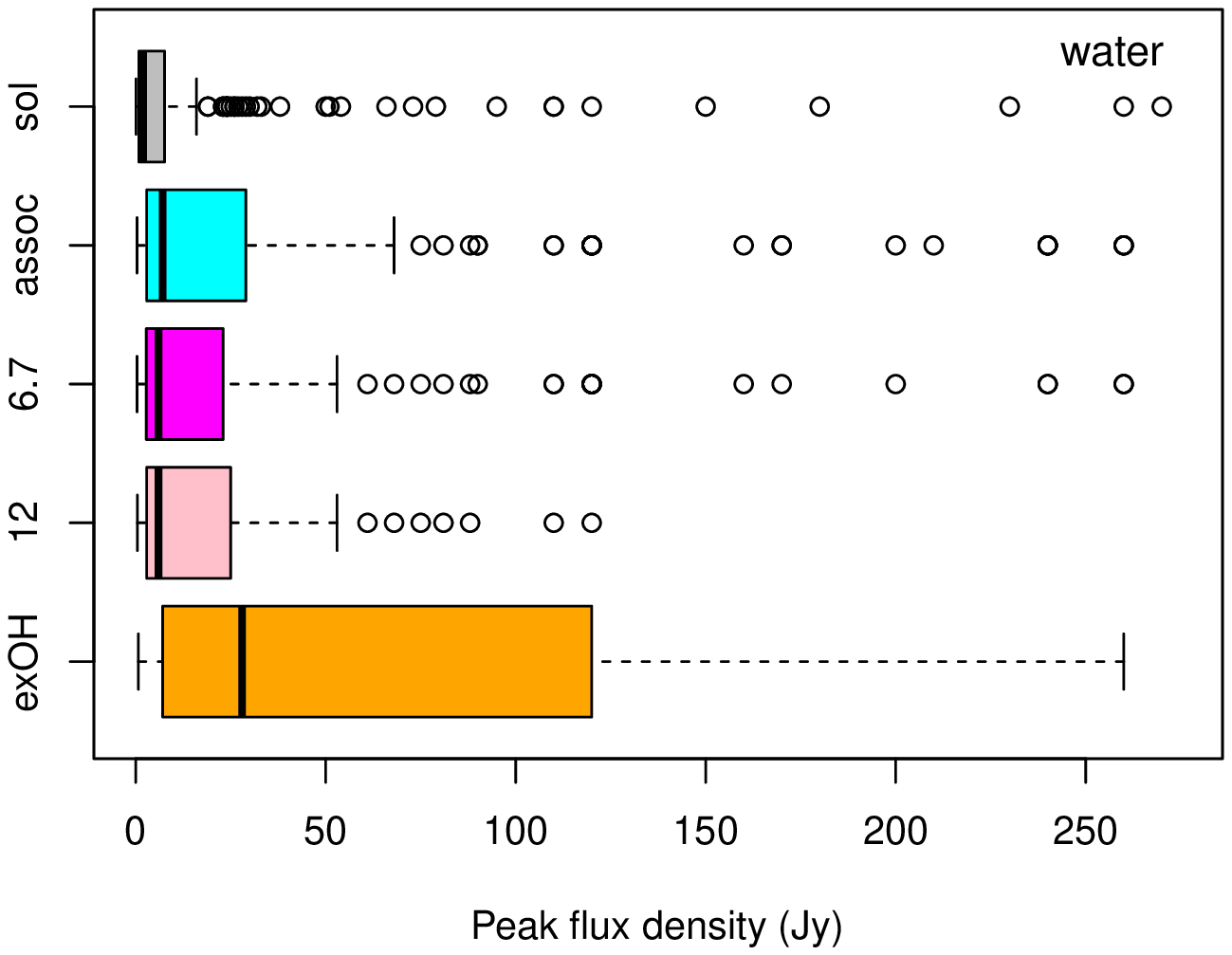,height=8cm,angle=270}
\caption{Peak flux density box plots of 6.7-GHz methanol masers (top left), 12.2GHz methanol masers (top right), excited-state OH masers (bottom left) and water masers (bottom right) in the categories of solitary (`sol'; grey), associated with one or more of the other transitions (`assoc'; cyan), associated with 6.7-GHz methanol (`6.7'; magenta), 12.2-GHz (`12'; pink), excited-state OH (`exOH'; orange) and water masers (`water'; blue). Note that in the case of 12.2-GHz methanol masers `assoc' includes those 12.2-GHz sources that are associated with either or both of excited-state OH or water maser emission since all are associated with 6.7-GHz methanol maser sources. All plots have been truncated in order to see details: the 6.7-GHz plot is missing nine sources, the 12.2-GHz plot is missing six sources, the excited-state OH plot is missing one source and the water plot is missing six sources. See Fig.~\ref{fig:vel_box} caption for a general explanation of box plots.}

\label{fig:flux_box}
\end{center}
\end{figure*}

Fig.~\ref{fig:flux_box} shows box plots of the peak flux density distribution of each of the four maser types in association categories of solitary (those masers not associated with any of the three other maser types), associated (those masers associated with one or more of the other maser types) and then associations with each of the maser types separately. In every case the solitary masers show the lowest peak flux densities and similarly for the 12.2-GHz sources (which are never solitary) the lowest peak flux density distribution are for those sources not associated with either excited-state OH masers or water maser emission. In each case the highest median and 75th percentile value is for those masers that are associated with excited-state OH emission and the second highest median value is for those sources that are associated with 12.2-GHz methanol maser emission.

\subsection{Distribution of 6.7- and 12.2-GHz luminosities}

\begin{figure}\vspace{-1.5cm}
\begin{center}
	\epsfig{figure=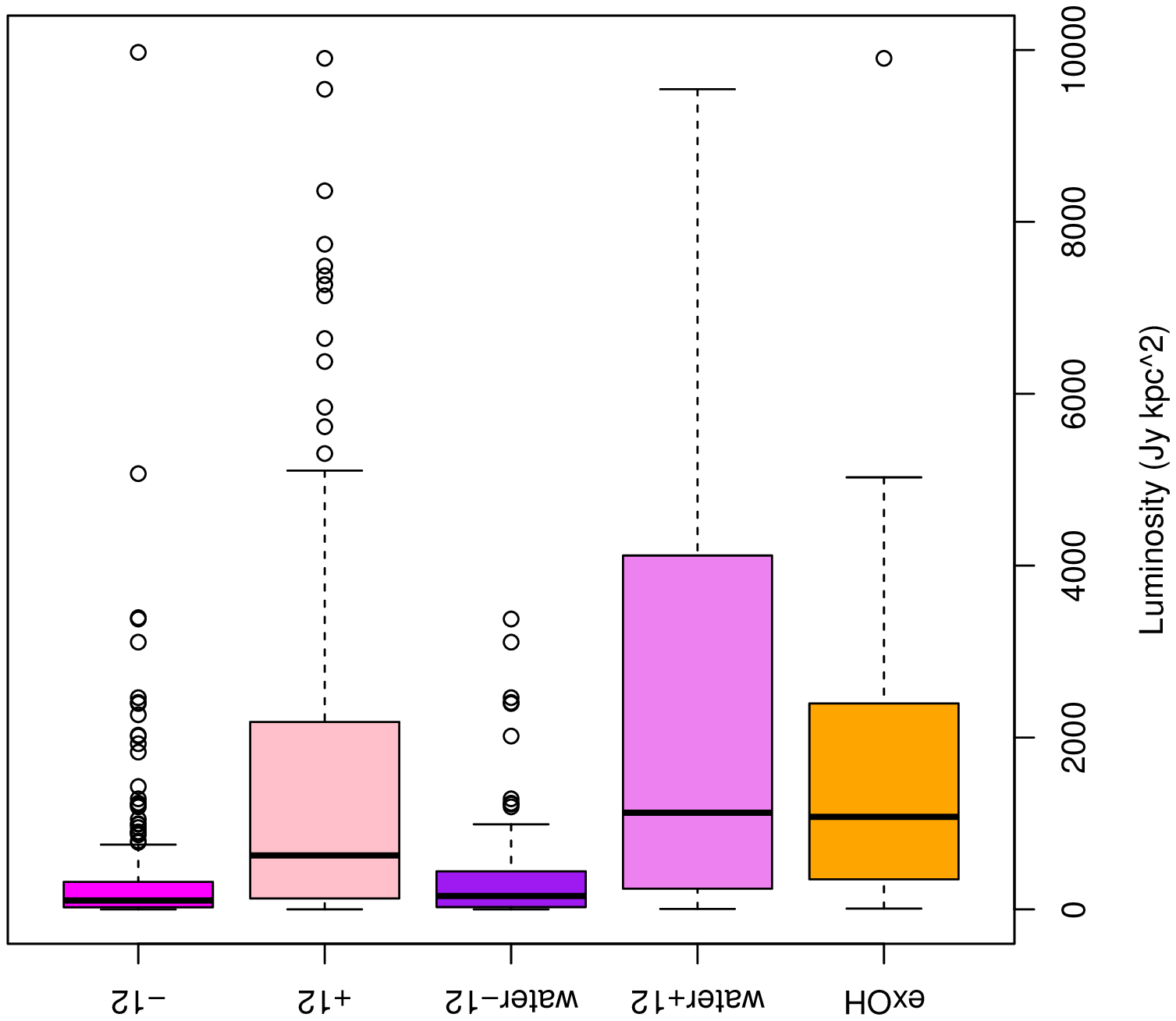,height=9cm,angle=270}
\caption{Peak luminosity of 6.7-GHz methanol masers with (+12; pink) and without (-12; magenta) accompanying 12.2-GHz methanol masers emission, with water but no 12.2-GHz (water-12; purple), with both 12-GHz and water (water+12; violet) and with excited-state OH maser emission (exOH; orange). Luminosities have been calculated using distances from \citet{GM11}, \citet{Green17} and references therein (where available). There are 16 6.7-GHz methanol masers with luminosities greater than 10000 Jy \kms and are therefore excluded from the plot (470 have luminosities less than 10000). See Fig.~\ref{fig:vel_box} caption for a general explanation of box plots.}
\label{fig:lum6.7}
\end{center}
\end{figure}

Fig.~\ref{fig:lum6.7} shows the luminosity of 6.7-GHz methanol masers in a number of different association categories. The categories of 6.7-GHz methanol masers with no accompanying 12.2-GHz emission and with accompanying water but no 12.2-GHz masers show the lowest luminosities. The distributions of 6.7-GHz methanol maser luminosity extends to much higher values for those sources that are associated with 12.2-GHz methanol masers and even higher still for sources with both 12.2-GHz and water maser emission. 6.7-GHz methanol maser emission with accompanying excited-state OH emission shows a distribution similar to those sources with 12.2-GHz.

\citet{Breen12stats} found evidence that 6.7-GHz methanol masers increase in luminosity as they evolve, suggesting that Fig.~\ref{fig:lum6.7} might imply the following evolutionary scenario: 6.7-GHz methanol masers devoid of 12.2-GHz methanol maser emission are the youngest and 6.7-GHz methanol masers showing 12.2-GHz methanol or excited-state OH masers are present at a slightly later phase of evolution. The presence of a water maser may signal a slightly later evolutionary phase.

\section{Galactic distribution of the different maser transitions}

\begin{figure}\vspace{-1.5cm}
\begin{center}
	\epsfig{figure=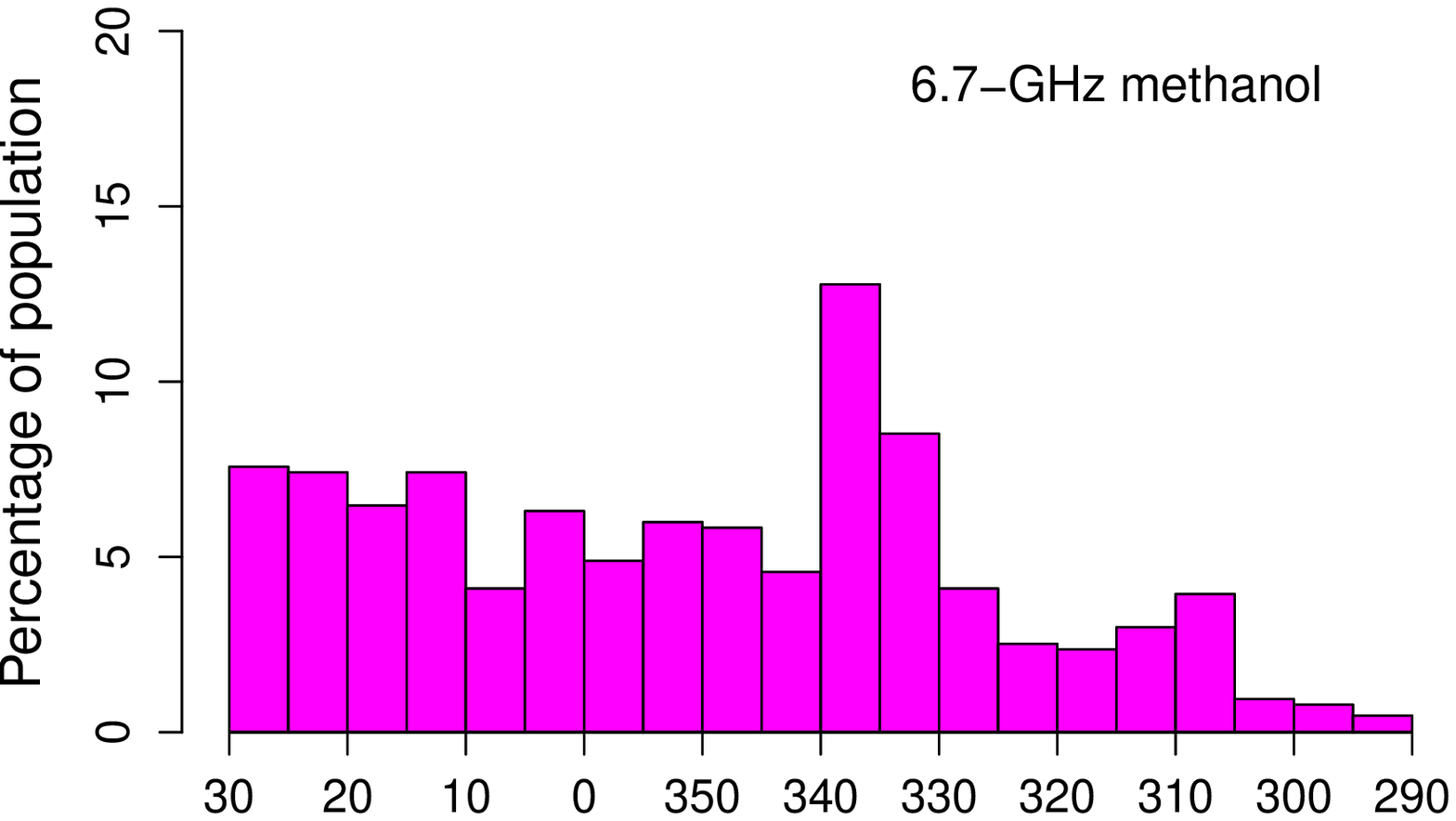,height=9cm,angle=270}\vspace{-1.5cm}
	\epsfig{figure=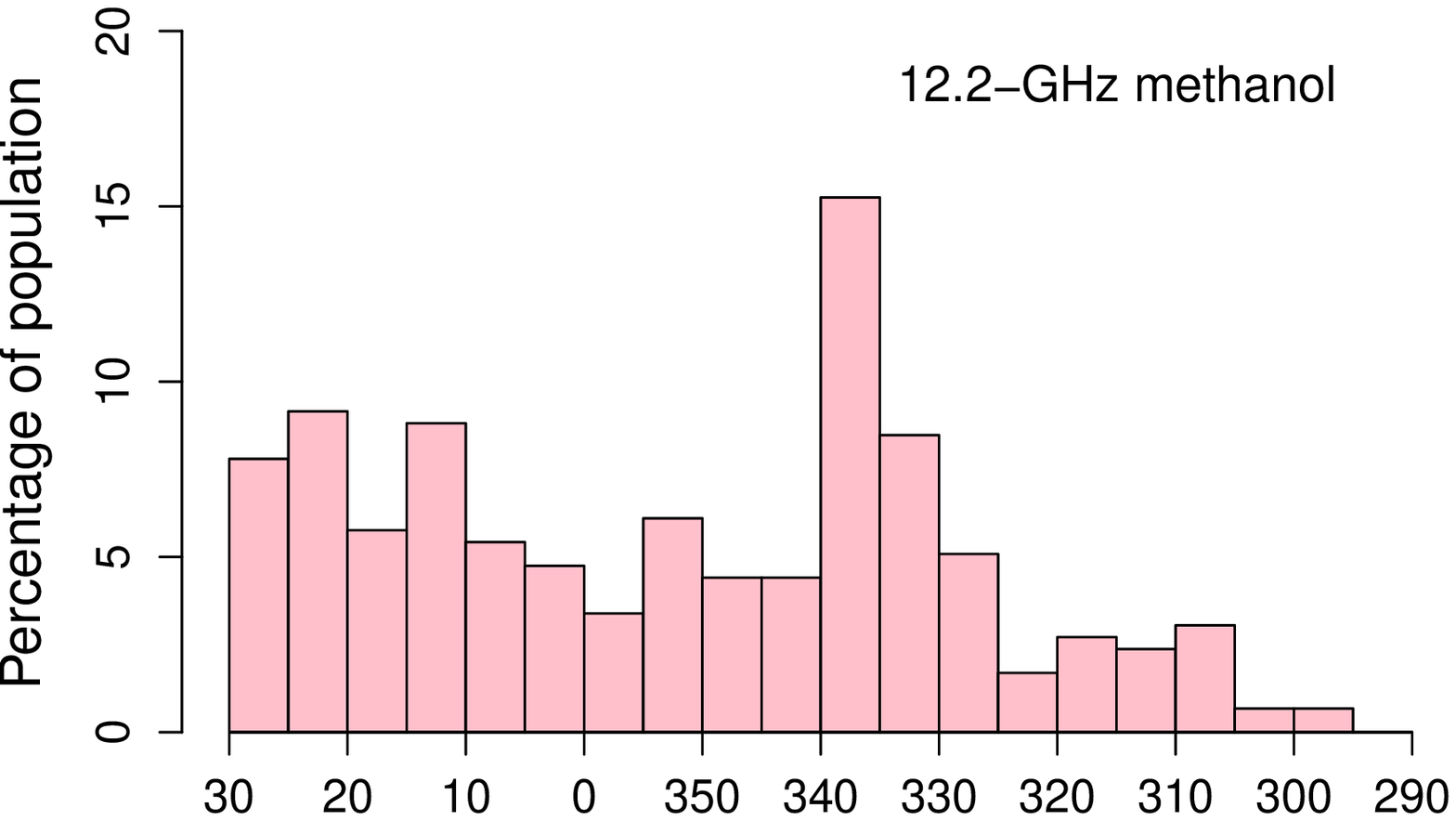,height=9cm,angle=270}\vspace{-1.5cm}
	\epsfig{figure=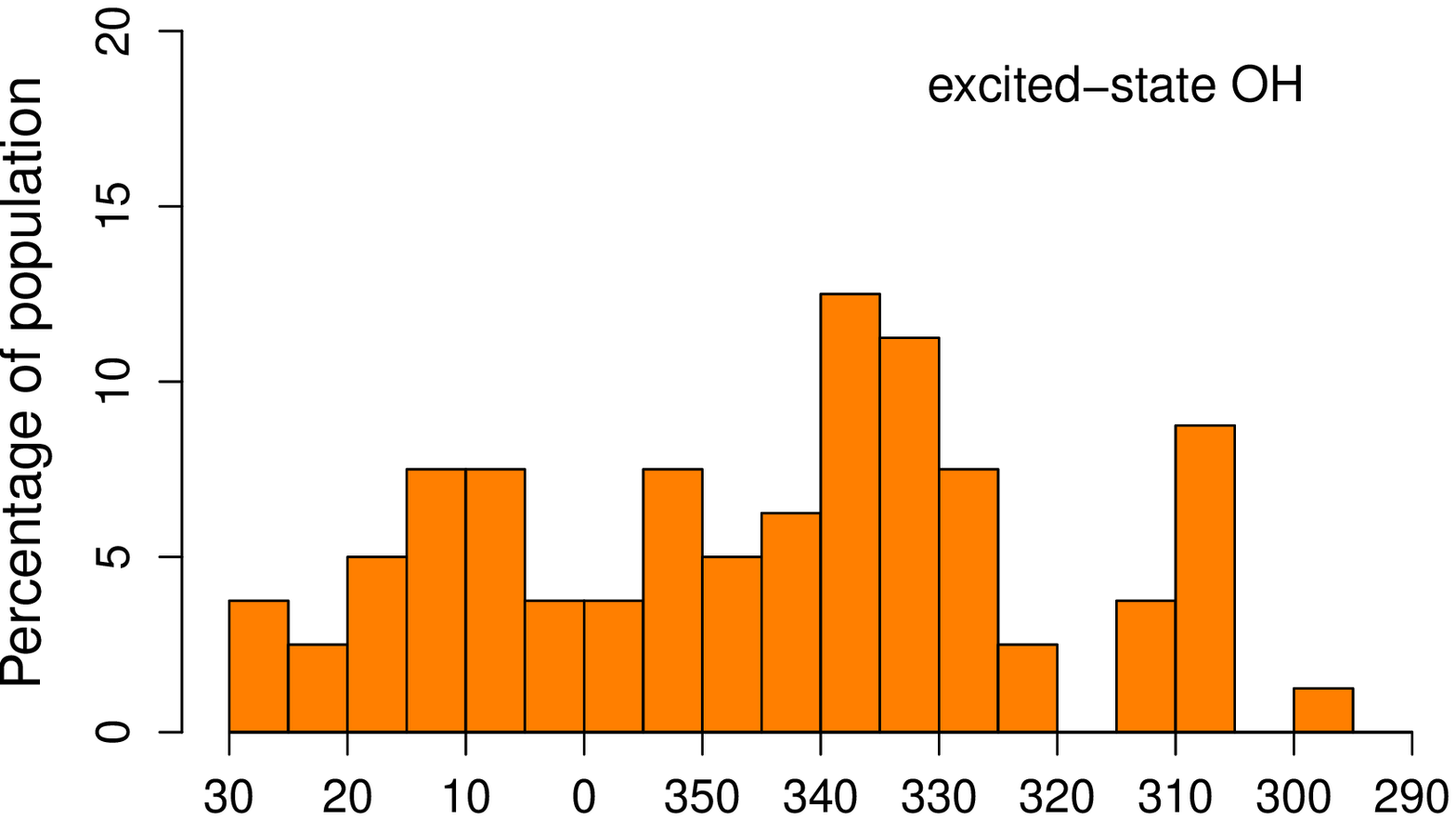,height=9cm,angle=270}\vspace{-1.5cm}
	\epsfig{figure=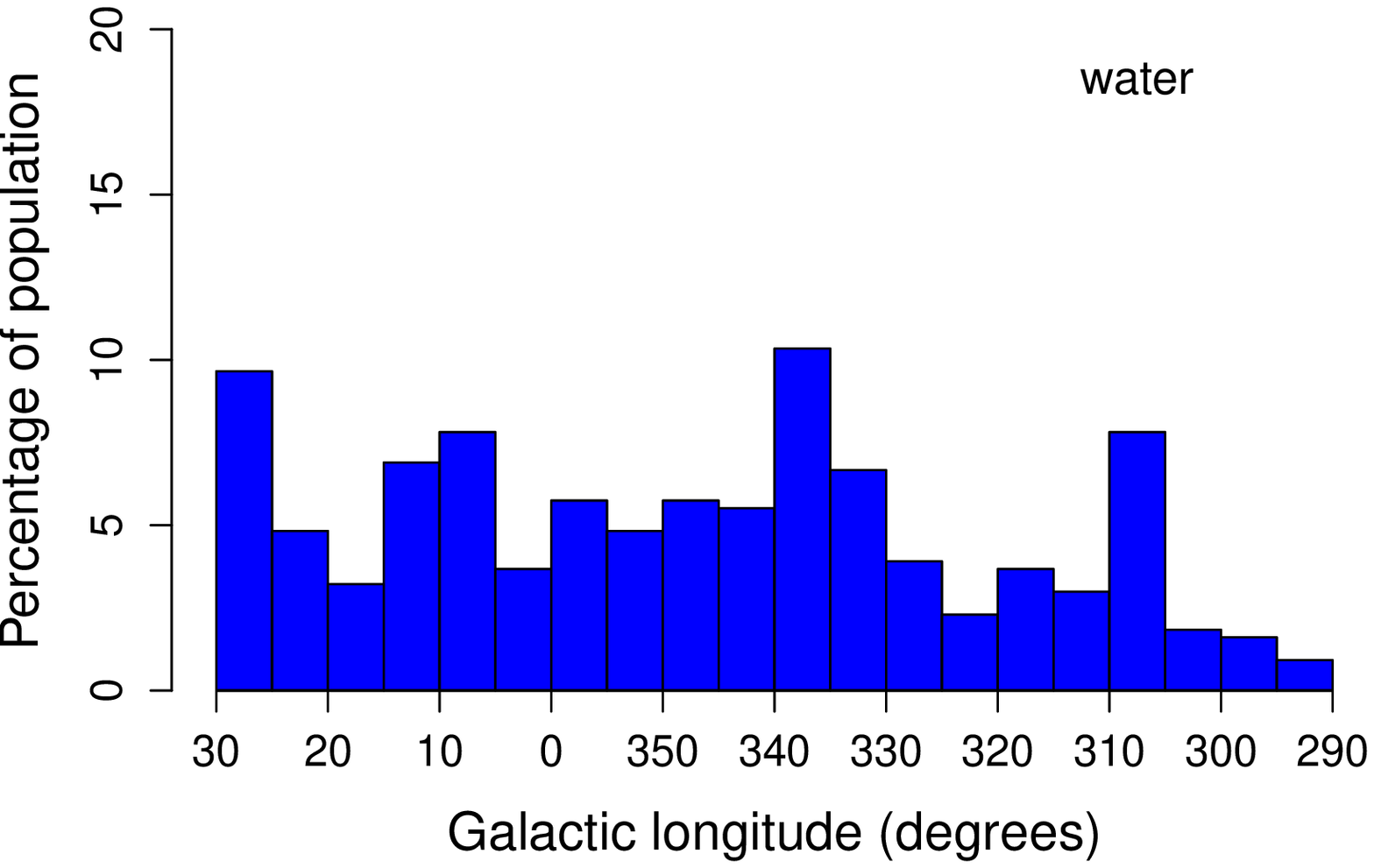,height=9cm,angle=270}
\caption{Histograms of the percentage of 6.7-GHz methanol, 12.2-GHz methanol, excited-state OH and water masers as a function of Galactic longitude.}
\label{fig:long_percentage}
\end{center}
\end{figure}

\begin{figure}\vspace{-1.5cm}
\begin{center}
	\epsfig{figure=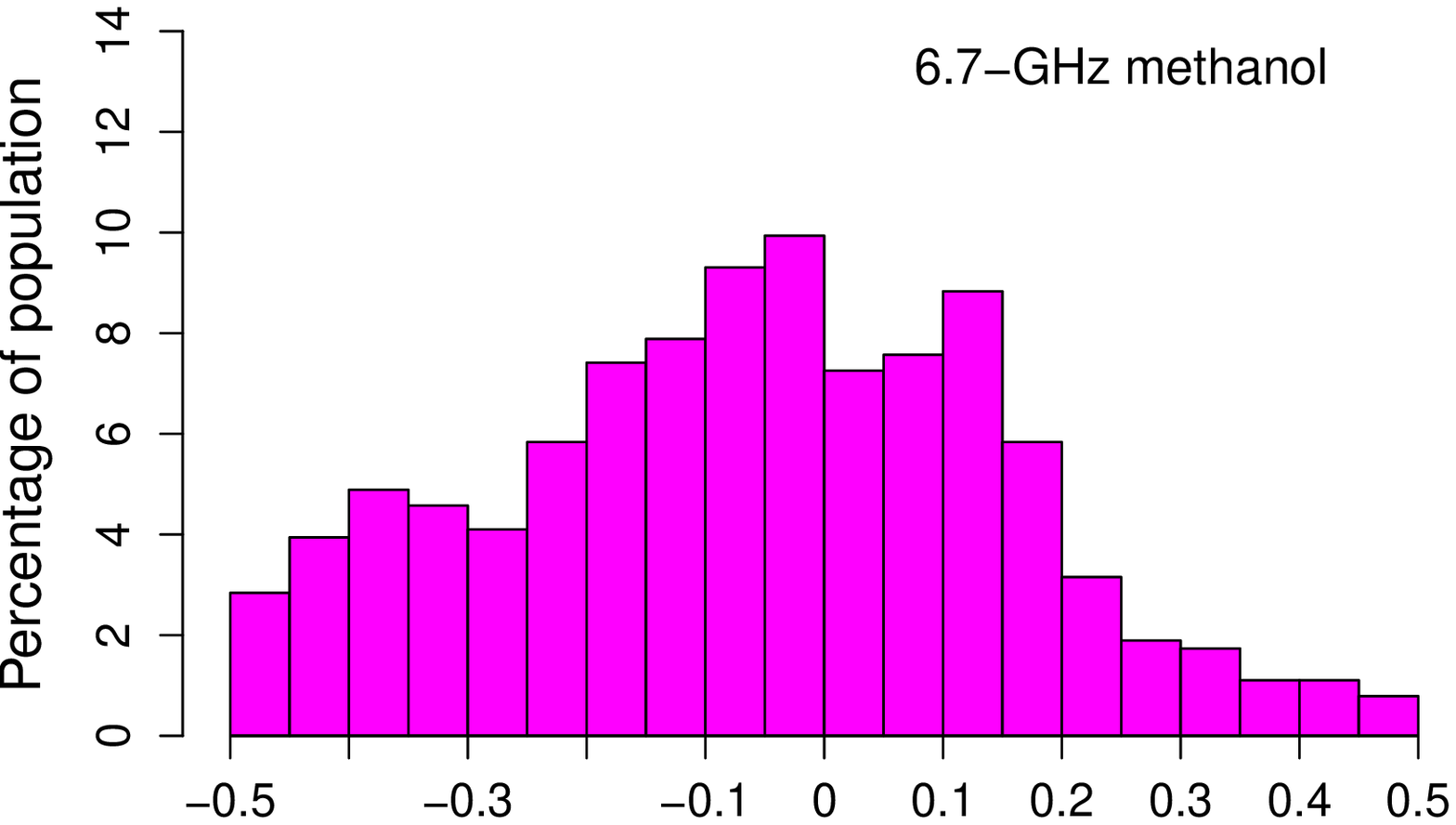,height=9cm,angle=270}\vspace{-1.5cm}
		\epsfig{figure=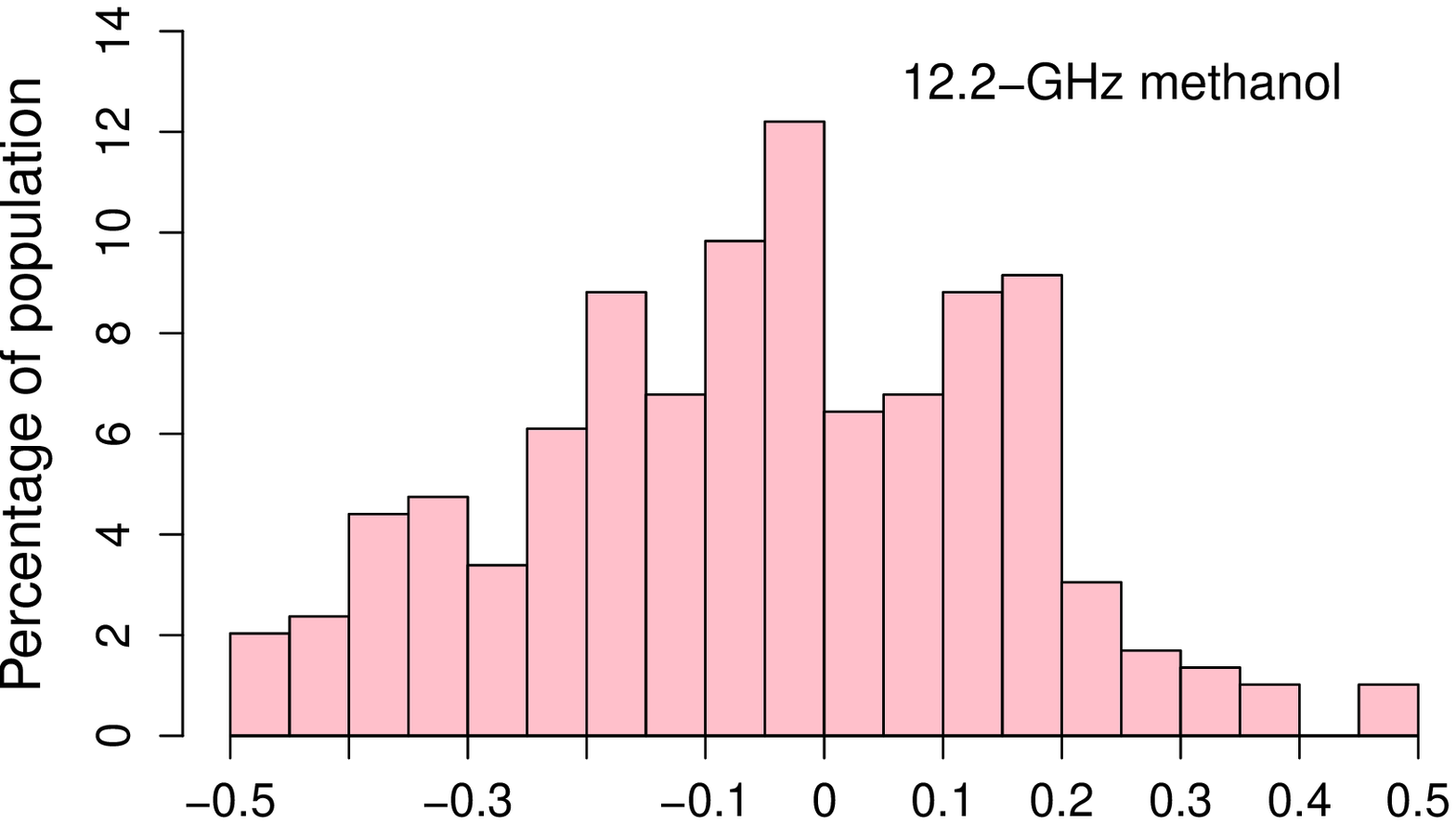,height=9cm,angle=270}\vspace{-1.5cm}
	\epsfig{figure=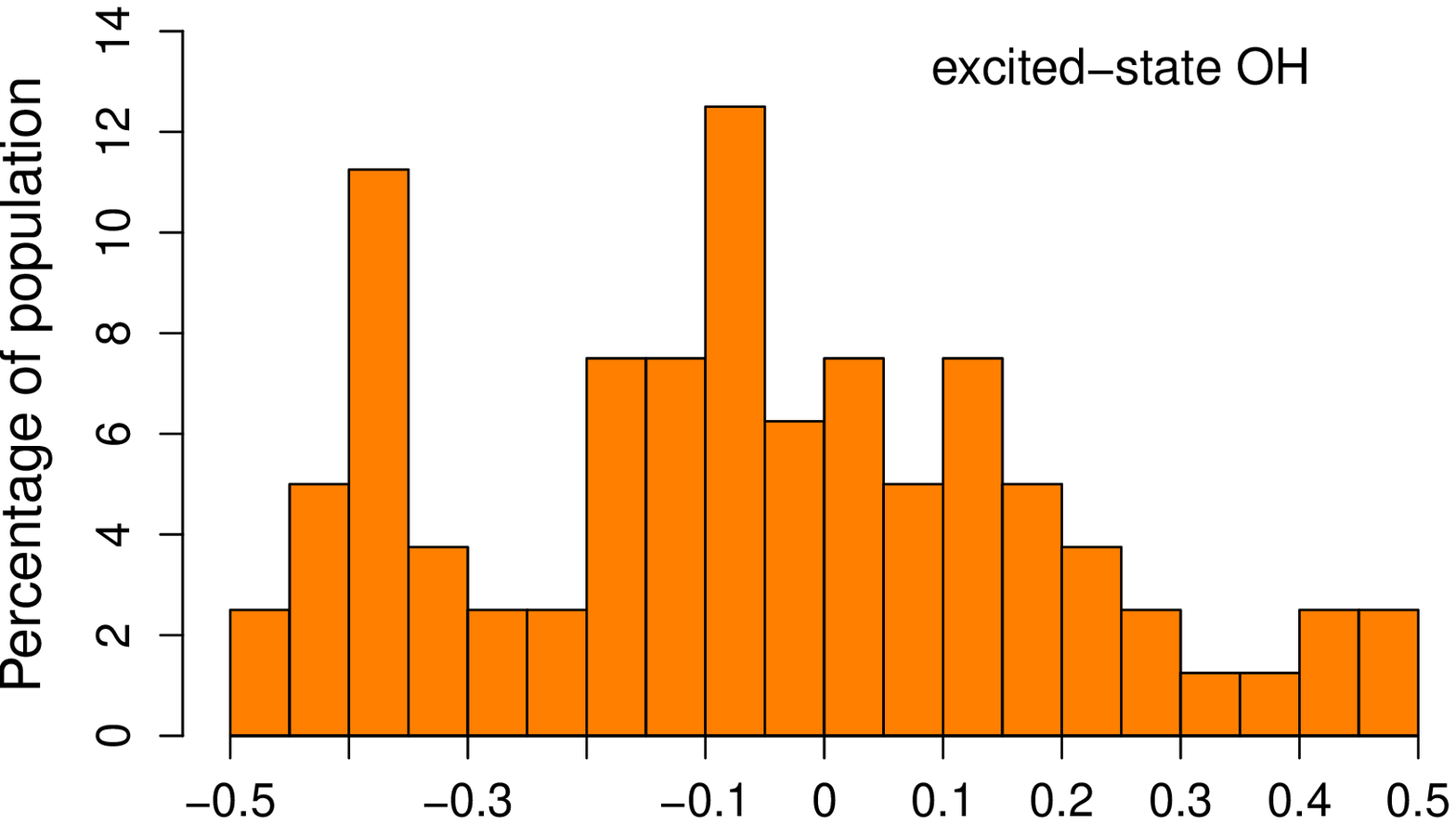,height=9cm,angle=270}\vspace{-1.5cm}
	\epsfig{figure=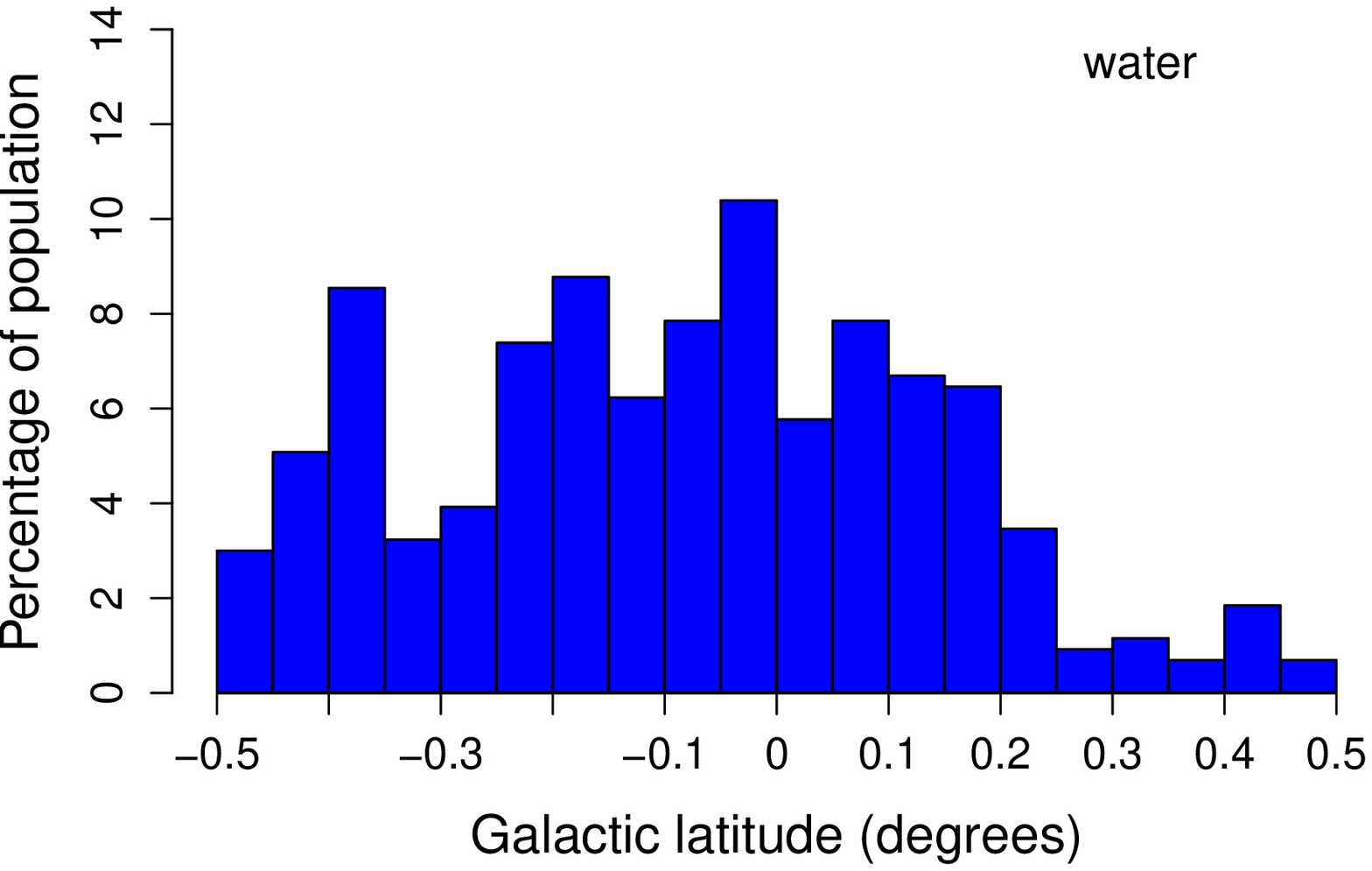,height=9cm,angle=270}
\caption{Histograms of the percentage of 6.7-GHz methanol, 12.2-GHz methanol, excited-state OH and water masers as a function of Galactic latitude. Note that there are only 433 of the star formation water masers plotted here since two of them fall slightly outside the latitude range.}
\label{fig:lat_percentage}
\end{center}
\end{figure}

The population distributions of each maser transition are plotted in a series of histograms in Fig.~\ref{fig:long_percentage} and \ref{fig:lat_percentage} as a function of Galactic longitude and latitude, respectively. The distributions are similar in all cases, and indeed K-S tests show that there is no statistically significant evidence that any of the maser populations are drawn from different distributions. However, we note that the G\,335 to G\,340 longitude range shows the greatest abundance of each of the four maser types and corresponds to the Perseus arm origin. Qualitatively it also appears that the two methanol maser distributions are very similar to each other, but slightly different from the water and excited-state OH distributions which are also much more similar to each other. Perhaps the most pronounced difference in the distributions is the relative overabundance of both excited-state OH and water masers in the latitude range $-$0.35$^{\circ}$ to $-$0.40$^{\circ}$ compared to the 6.7- and 12.2-GHz methanol masers. A similar feature is seen in the longitude distributions between longitudes G\,305$^{\circ}$ and G\,310$^{\circ}$, approximately corresponding to the Crux-Scutum arm tangent, but does not correspond to the same sources in the latitude range $-$0.35$^{\circ}$ to $-$0.40$^{\circ}$. 

Fig.~\ref{fig:lat_percentage} shows some evidence that the 6.7- and 12.2-GHz methanol masers are much more tightly constrained to lower latitudes than either the water masers or the excited-state OH masers. Although broader latitude coverage would be needed to confirm this, it is consistent with the idea that a significant fraction of the methanol masers are tracing a generally earlier phase of star formation than the other maser types. In the case of the water masers, a broader latitude distribution might be further contributed to by their association with a broader mass range.

\citet{Green11} identified several regions of 6.7-GHz methanol maser density enhancements, corresponding to signifiant structures within the Galaxy (such as arm origins). Within the HOPS region, there are 224 6.7-GHz methanol masers that are likely to be associated with these structures \citep[see table 1 of][for longitude, latitude and velocity ranges corresponding to each region of enhancement]{Green11}. We find that the 6.7-GHz methanol sources that are associated with these structures have similar association rates with 12.2-GHz, excited-state OH and water masers to the entire region (47.8, 6.3, 27.2 per cent association rates, respectively compared to 45.9, 7.1 and 29.8 per cent for the 6.7-GHz methanol masers that are not associated with major structures). We have also compared the association rates of 12.2-GHz, excited-state OH and water maser emission associated with the 45 6.7-GHz methanol masers from the MMB survey found to be associated with the 3-kpc arms \citep{GreenMMB10}. We find that the association rates with 12.2-GHz, excited-state OH and water masers are all lower, but given the small numbers, essentially equivalent to the larger survey region.

Fig.~\ref{fig:lat_percentage} shows that the latitude distribution of the different maser species all peak at negative values. This is a result that has been found in an increasing large number of Galactic plane surveys; both the Bolocam Galactic Plane Survey \citep[BGPS;][]{BGPS10} and the APEX Telescope Large Area Survey of the GALaxy \citep[ATLASGAL;][]{ATLASGAL09,Beuther12,Contreras13} surveys of dust continuum emission found a latitude distribution peak at $-$0.09$^{\circ}$. \citet{Reed06} measured the location of the Sun to be $\sim$20~pc above the Galactic plane, which will skew sources within the plane of the Galaxy to negative values. Indeed, both \citet{ATLASGAL09} and \citet{Contreras13} attribute the excursion of their sources from the Galactic plane to the location of the Sun above the plane.

The 6.7-GHz methanol masers have a mean latitude of $-$0.06$^{\circ}$$\pm$0.009$^{\circ}$, the excited-state OH masers have a mean latitude of $-$0.06$^{\circ}$$\pm$0.03$^{\circ}$ and the water masers have a mean latitude of $-$0.09$^{\circ}$$\pm$0.01$^{\circ}$. The offset between the dust continuum surveys and the 6.7-GHz methanol masers is likely because the MMB is much more sensitive to sources on the far side of the Galaxy which do not show the same offset from the Galactic plane. In fact, the latitude distribution to 6.7-GHz sources located at distances farther than 8~kpc shows a mean latitude that is consistent with zero ($-$0.01$^{\circ}$$\pm$0.01$^{\circ}$). Splitting these distant sources into two groups based on quadrants, results in mean latitude values of $-$0.04$^{\circ}$$\pm$0.02$^{\circ}$ for the first quadrant sources and 0.004$^{\circ}$$\pm$0.02$^{\circ}$ for the fourth quadrant, consistent with the expectations of the Galactic warp.

\section{Comparing the locations of masers with temperature data}

\subsection{Temperatures derived from HOPS ammonia observations}

Alongside the water maser observations, HOPS simultaneously observed a number of ammonia transitions \citep{Walsh11}. \citet{Purcell12} presented a catalogue of the ammonia (1,1) and (2,2) from the HOPS survey, which has a mean sensitivity of 0.20~K and catalogues 669 dense molecular clouds. Using these data, \citet{Longmore17} investigated the properties of the dense gas and made some comparisons between those sources with and without 6.7-GHz methanol masers and HOPS water masers. \citet{Longmore17} were able to derive kinetic temperatures for 64 ammonia sources and they looked at maser associations and found that those sources with either 6.7-GHz methanol masers or water masers were warmer than those without and also had larger linewidths.

We have compared the ammonia sources associated with any combination of  6.7-, 12.2-GHz methanol, excited-state OH or water maser emission and find that those associated with masers have a higher average and median temperature (28.5~K and 24~K) compared to those without any maser emission (25.5~K and 22~K), although due the significant overlap in the temperatures and the small sample size, this difference is not statistically significant. Since there are only 64 sources in the \citet{Longmore17} sample, we found that splitting the sources into association categories based on the different combinations of detected masers yielded no meaningful results. 

\subsection{Far-infrared dust temperatures}

\citet{Guzman15} calculated far-infrared dust temperatures using Hi-Gal (Herschel Infrared Galactic Plane Survey) and ATLASGAL (APEX Telescope Large Area Survey of the Galaxy) data, covering wavelengths between 160- and 870-$\mu$m. While their work investigated the dust temperatures of $\sim$3000 high-mass star forming clumps from the MALT90 survey \citep{Jackson13}, their temperature maps cover the full range of HOPS and can therefore be used to compare with the different combinations of maser species. The resolution of the temperature maps is limited by that of the 500-$\mu$m maps to $\sim$35\arcsec.

\begin{figure}\vspace{-1.5cm}
\begin{center}
	\epsfig{figure=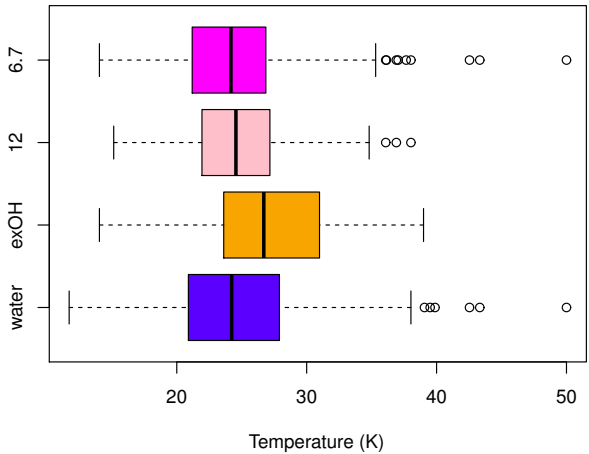,height=9cm,angle=270}
	
\caption{Far infrared dust temperatures of regions exhibiting 6.7-GHz methanol masers (6.7; magenta), 12.2-GHz methanol masers (12; pink), excited-state OH masers (exOH; orange) and water masers (water; blue). See Fig.~\ref{fig:vel_box} caption for a general explanation of box plots.}
\label{fig:temp_box}
\end{center}
\end{figure}

Fig.~\ref{fig:temp_box} shows the far-infrared temperature distribution of regions associated with each of the four types of masers. Although there is a lot of overlap in each of the groups - both because some sources are present in more than one of the categories and also because temperatures have been averaged across significant regions of sky - there are statistically significant differences between the groups. According to a Wilcoxon rank-sum test there are statistically significant differences between the far-infrared dust temperatures at the locations of the excited-state OH masers and each of the other masers types (p-values all smaller than 0.002) and a similar result is found by running a K-S test, which shows that the distributions are significantly different (p-values all better than 0.01). None of the other maser categories show temperatures that are statistically significantly different from any of the other groups but this is largely due to some limitations caused by overlap in the datasets (especially since all of the 12.2-GHz sources are also present in the 6.7-GHz source category, for example). If we compare the temperatures of regions associated with just 6.7-GHz methanol masers to the regions that have both 6.7- and 12.2-GHz emission we find that the temperatures at the locations of 12.2-GHz masers are significantly warmer (see Table~\ref{tab:12temps}) than those without (Wilcoxon rank-sum gives a p-value of 0.05 and a K-S test shows that the distributions are significantly different, with a p-value of 0.02).

Class~II methanol masers have provided some of the clearest maser evolutionary trends to date, and is thought to be a result of the close connection between the masers and the protostar ensured by radiative pumping. In this evolutionary scenario, 6.7-GHz methanol masers appear significantly earlier than their 12.2-GHz counterparts \citep{Breen10b} and, other, rarer class~II methanol masers at 37.7-GHz methanol masers trace a short-lived phase, present just prior to the cessation of the class~II methanol maser phase \citep{Ellingsen11,Ellingsen13}. That the far-infrared dust temperatures are higher for 6.7-GHz masers associated with 12.2-GHz masers than 6.7-GHz masers devoid of accompanying 12.2-GHz emission, provides an independent corroboration that 12.2-GHz methanol masers are present at a later evolutionary phase. It is, of course, possible that mass is a confounding variable here, but given the resolution of 35\arcsec\ (corresponding to a linear resolution of 0.85~pc at a distance of 5~kpc), evolution is a much more likely culprit given that it will take a significant time to heat the surrounding medium enough to result in a spatially averaged temperature difference.

\begin{table}\footnotesize
 \caption{Mean and median dust temperatures in a number of association categories; in the first section 6.7-GHz methanol masers with and without associated 12.2-GHz methanol maser emission are shown. The standard deviation is given in column four. The second and third vertically separated sections correspond to the data shown in Fig~\ref{fig:6.7_dust_temp}.}
  \begin{tabular}{lccc} \hline
 \multicolumn{1}{c}{Association} &\multicolumn{1}{c}{Mean} &\multicolumn{1}{c}{Median} & \multicolumn{1}{c}{Standard}  \\
 \multicolumn{1}{c}{Category}&\multicolumn{1}{c}{Temp (K)} &\multicolumn{1}{c}{Temp (K)}  &\multicolumn{1}{c}{deviation}\\ \hline
6.7-GHz with 12.2-GHz	& 24.7	&	24.6 &	4.2\\ 
6.7-GHz no 12.2-GHz	&	24.2	& 23.8 &	4.8 \\ \hline

solitary 6.7-GHz & 23.6	& 23.0 &	4.2\\ 
associated 6.7-GHz & 24.9	& 24.8 &	4.6\\ 
6.7-GHz with 12.2-GHz	& 24.7	&	24.6 & 4.2\\ 
6.7-GHz with water &25.4 &25.2 &	4.8\\ 
6.7-GHz with exOH & 26.7 & 27.0 &	5.0\\ 
\hline

solitary water	& 23.8	&	23.0 & 5.8\\
associated water	&	25.6	& 25.2 & 5.0\\
water with 6.7-GHz&25.4 &25.2 &	4.8\\ 
water with 12-GHz		&	25.5	&	25.8 & 3.8	\\
water with exOH	&	26.9	&	26.7 & 6.3\\ \hline
\end{tabular}\label{tab:12temps}
\end{table}

\begin{figure}
\begin{center}\vspace{-1.5cm}
	\epsfig{figure=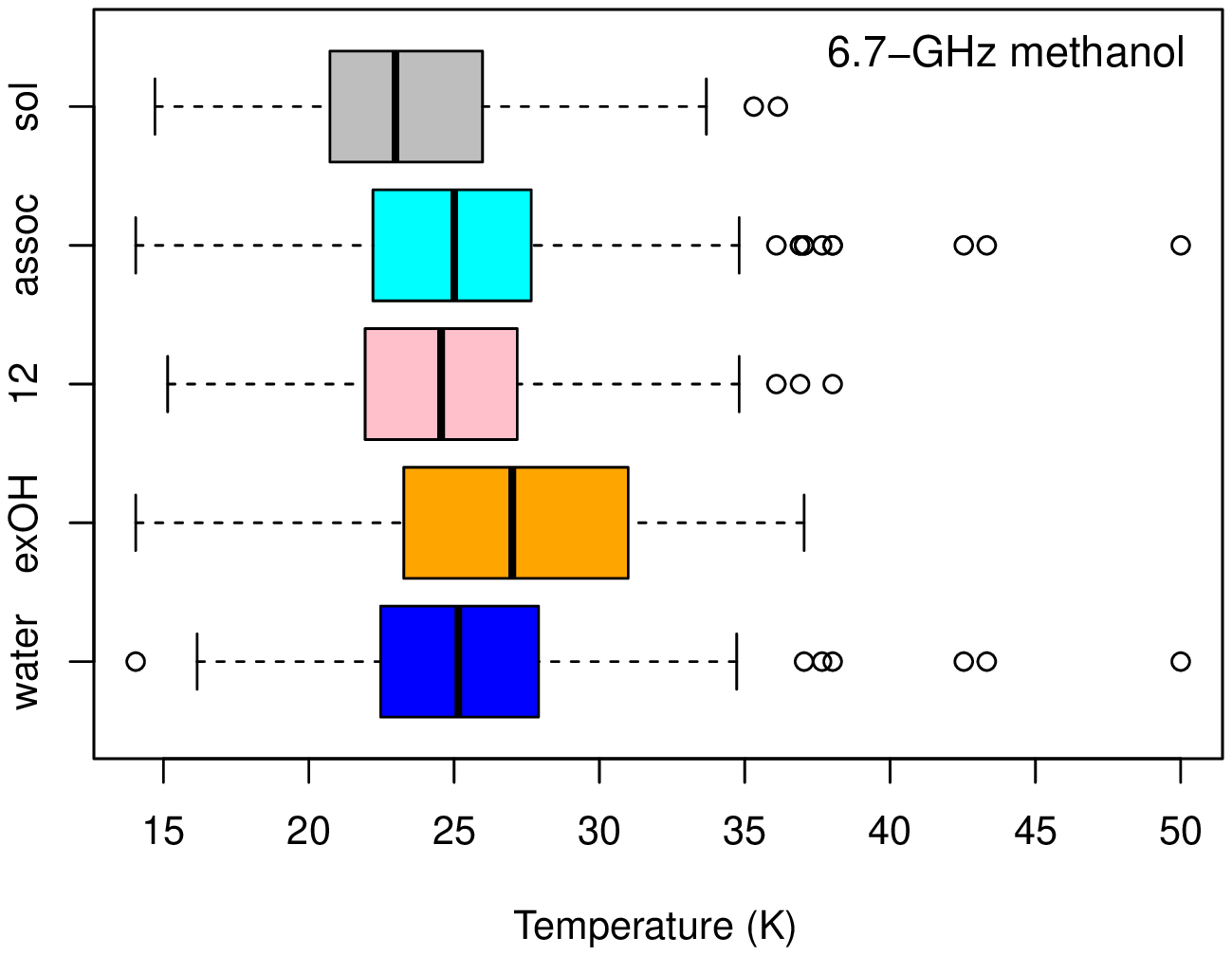,height=9cm,angle=270}\vspace{-1cm}
		\epsfig{figure=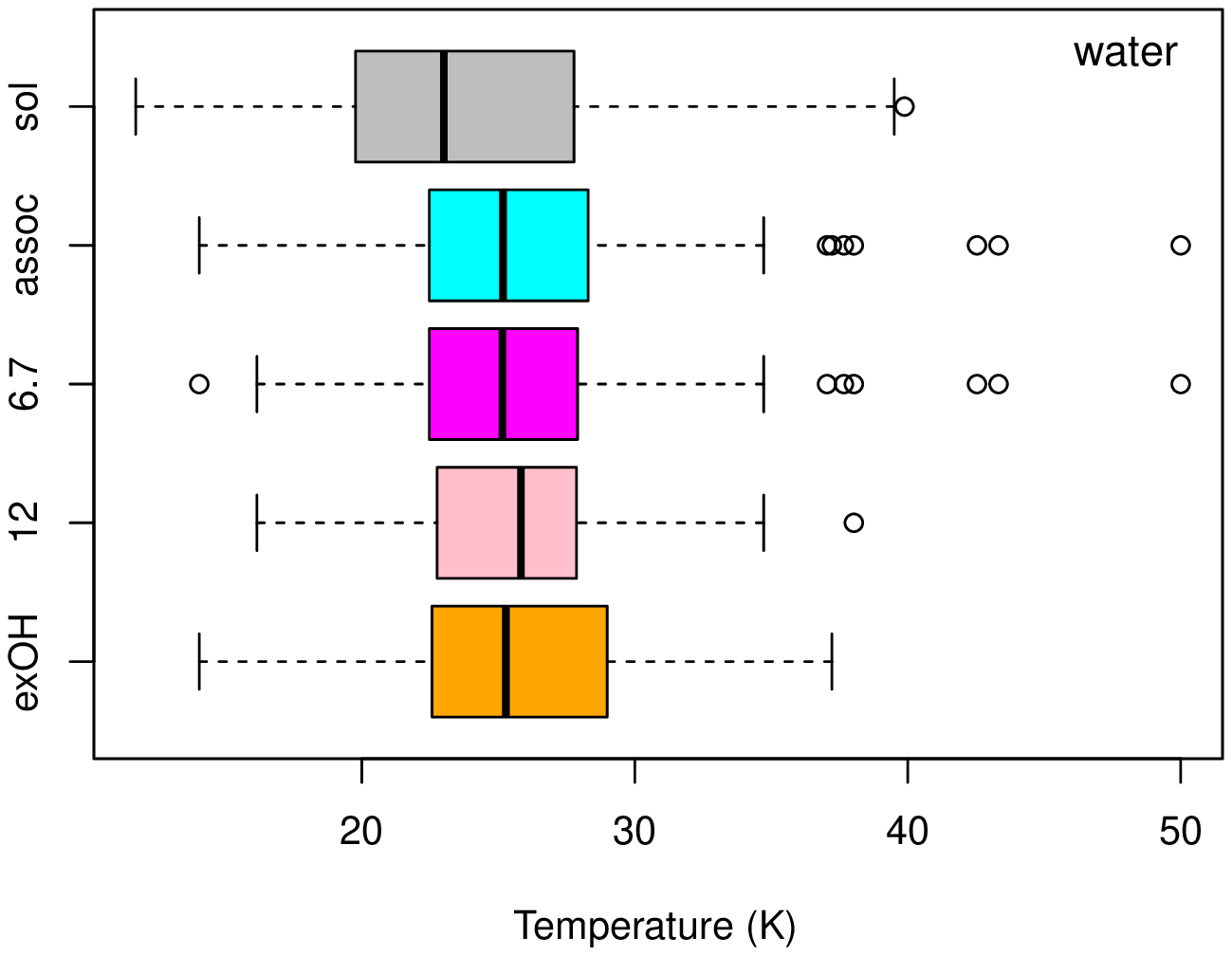,height=9cm,angle=270}
\caption{Temperatures of the regions showing (top) just 6.7-GHz methanol maser emission (`sol'; grey), as well as those that are associated with one or more maser type (`assoc'; cyan) and associations with 12.2-GHz methanol masers (`12'; pink), excited-state OH masers (`exOH'; orange) and water masers (`water'; blue) and (bottom) just water maser emission (`sol'; grey), as well as those that are associated with one or more maser type (`assoc'; cyan) and associations with 6.7-GHz methanol masers (`6.7'; magenta), 12.2-GHz methanol masers (`12'; pink) and excited-state OH masers (`exOH'; orange). See Fig.~\ref{fig:vel_box} caption for a general explanation of box plots.}
\label{fig:6.7_dust_temp}
\end{center}
\end{figure}

Fig.~\ref{fig:6.7_dust_temp} shows the temperatures of sources exhibiting 6.7-GHz methanol and water masers, respectively, in each of the association categories. The distribution of solitary sources are skewed towards lower temperatures in both cases, but interestingly the temperature of the solitary 6.7-GHz masers are not drawn from the same population as the solitary water masers (p-value from K-S test 0.03). Table~\ref{tab:12temps} shows that there is little difference in the median and mean values of these categories but Fig.~\ref{fig:6.7_dust_temp} shows that the temperatures of regions associated with solitary water masers cover a broader range than those associated with solitary 6.7-GHz masers (interquartile ranges of 8.0~K and 5.3~K, respectively).

In each case, the association categories show statistically significant differences in temperature compared to solitary water or methanol masers (resultant p-values from K-S tests are all smaller than 0.002, except for water masers that also show excited-state OH maser emission which has a p-value of 0.048). Additionally we find that the association categories of 6.7-GHz with accompanying 12.2-GHz are drawn from a different population than the 6.7-GHz with excited-state OH masers (p-value 0.015).

\citet{Breen12stats} found that both the luminosity and the velocity ranges of sources increased with evolution, suggesting in the latter case that this was a result of either increased internal motions or that the volume of gas conducive to the maser action increased as the sources evolved. We have compared the dust temperatures of the 6.7-GHz methanol maser sources with the top 20 per cent of velocity ranges to the rest of the sample and find that the sources with the highest velocity ranges have both higher average (24.8~K compared with 24.3~K) and median (24.8~K compared with 24.0~K) temperatures. 

We find a similar result if we look at 6.7-GHz methanol maser luminosity instead (which is expected given the relationship found between maser velocity range and luminosity) - the most luminous 20 per cent of maser sources (that have temperature data) are warmer, on average (25.1 compared with 24.1~K with medians of 25.1 and 23.8~K). While these results are all self consistent, none of the comparisons made between velocity ranges or luminosities and dust temperatures are statically significant, meaning that average and median values are higher, but there is no statistical evidence there is a population difference. 

Since the resolution of the temperature and column density maps is relatively low (35 arcsec) compared to individual young high-mass stars, the small temperature difference is most likely representing a much larger temperature difference that has been spatially averaged, diluting the true differences. The fact that there is no tight one-to-one correlation in any of these of these comparisons is likely a consequence of this.

\subsubsection{Comparison with maser models}

Comparisons between expectations from maser modelling and measured physical conditions with respect to maser occurrence has the potential to significantly increase our understanding of the masers themselves, as well as the regions in which they are located. That we find statistically different dust temperatures between sources showing excited-state OH masers compared to the other maser types, as well as 6.7-GHz methanol masers with and without accompanying 12.2-GHz methanol masers, gives us a unique opportunity to compare with maser models \citep[e.g.][]{Cragg02,Cragg05}. 

\citet{Cragg05} presented maser modelling of class II methanol masers and their figs~2 and 3 show how maser brightness varies with a number of parameters, including dust temperature, which shows a sharp increase in maser brightness at a higher dust temperature for 12.2-GHz methanol masers compared to 6.7-GHz methanol masers. Their findings are naively consistent with our results, showing that sources associated with both 6.7- and 12.2-GHz methanol masers have slightly warmer dust temperatures, however, there are a number of complex factors at play, including the fact that these parameter plots show how dust temperature varies as other parameters remain fixed, and the fact that our dust temperatures represent the average of a significant region.

\citet{Cragg02} considered the characteristics of methanol and OH maser emission in their models, including the 6.7 and 12.2-GHz transitions of methanol and the 6035-MHz transition of OH. They found that 6.7-GHz methanol masers appear at dust temperatures of $\sim$100~K and then decline as the gas temperature increases to that of the dust temperature, while the 6035-MHz excited state OH masers was quenched at gas temperatures that surpassed 70~K but were independent of dust temperature. It is somewhat difficult to reconcile these predictions with our results, given that they are largely based on gas temperature, which our low resolution dust temperatures can shed little light on.

\section{Summary}

We have compared the occurrence of 6.7-GHz and 12.2-GHz methanol masers with 6035-MHz excited-state OH and 22-GHz water masers within the 100 square degree region common to HOPS and the MMB survey. We find that there are 634 6.7-GHz methanol masers, 435 water masers, 295 12.2-GHz methanol masers and 80 excited state OH masers within the survey range. The water maser population is effected by the lower sensitivity of HOPS and we expect that a water maser survey that had a similar sensitivity to the MMB were conducted, the number would be significantly increased, perhaps to a level that exceeded the 6.7-GHz methanol maser population. 

Water masers show the highest median velocity range (9.8~\kmsns), followed by 6.7-GHz methanol masers (6.1~\kmsns), excited-state OH (4.3~\kmsns) and 12.2-GHz methanol masers (1.9~\kmsns). We additionally find that the median velocity range of each of the maser types is lower when considering only sources that are not associated with other maser types. More than one third of the solitary water masers exhibit only one spectral feature. 

The luminosity of 6.7-GHz methanol maser emission is lowest when these masers are either solitary, or only associated with water maser emission. 6.7-GHz methanol masers associated with 12.2-GHz emission or excited-state OH maser emission are much more luminous (whether or not they are also accompanied by water maser emission).

We find that 89 per cent of water masers exhibit their peak emission within $\pm$10~\kms of the 6.7-GHz methanol maser central velocity (which gives a reliable indication of the systemic velocity) and that water maser peak velocity is, in general, close to the systemic velocity than their central velocity. Excited-state OH masers generally have velocities much closer to the systemic velocity - 86 per cent within $\pm$5~\kms of the 6.7-GHz methanol maser central velocity. All six of the excited-state OH masers with greater velocity separations show redshifted velocities.

Comparison of the far-infrared dust temperatures of sources exhibiting each of the maser types shows that those exhibiting excited-state OH masers are statistically significantly warmer. Since sources can exhibit more than one of the four types of masers (resulting in some sources being present in each of the associated categories) we also looked explicitly at a number of additional categories. In each case, sources exhibiting solitary water or solitary 6.7-GHz methanol masers have lower far-infrared dust temperatures than sources exhibiting additional types of maser emission. We also found that the far-infrared dust temperatures associated with sources exhibiting both 6.7-GHz and 12.2-GHz sources were significantly warmer than those only showing 6.7-GHz maser emission. These results provide independent support to the idea that different masers are present at different stages in the evolution of high-mass star formation regions.

\section*{Acknowledgments}

We thank A. E. Guzm\'an for providing us with far-infrared dust temperature maps of the Galactic plane. The Parkes telescope, the Australia Telescope Compact Array and Mopra telescope are part of the Australia Telescope which is funded by the Commonwealth of Australia for operation as a National Facility managed by CSIRO. Financial support for this work was provided by the Australian
Research Council. This research has made use of: NASA's Astrophysics
Data System Abstract Service; and  the SIMBAD data base, operated at CDS, Strasbourg,
France. 

\end{document}